\documentclass[12pt]{iopart}

\usepackage{iopams}
\usepackage{graphicx}
\usepackage{cite}
\usepackage{color}
\usepackage{multirow}
\usepackage[section]{placeins}
\usepackage{amssymb}

\begin{document}

\setlength{\parindent}{24pt}

\title[Characteristics of a bipolar pulsed discharge]{Modeling characterisation of a bipolar pulsed discharge}

\author{Zolt\'an Donk\'o$^{1,2}$, Lenka Zaji\v{c}kov\'a$^3$, Satoshi Sugimoto$^2$, Anjar Anggraini Harumningtyas$^{2,4}$, Satoshi Hamaguchi$^2$}

\address{$^1$Institute for Solid State Physics and Optics, Wigner Research Centre for Physics, 1121 Budapest, Konkoly-Thege Mikl\'os str. 29-33, Hungary\\
$^2$Center for Atomic and Molecular Technologies,
Graduate School of Engineering, Osaka University,
2-1 Yamadaoka, Suita, Osaka 565-0871, Japan\\
$^3$Masaryk University, Brno, Czech Republic\\
$^4$Center for Accelerator Science and Technology, National Nuclear Energy Agency of Indonesia (BATAN)}

\ead{donko.zoltan@wigner.hu}

\begin{abstract}
We apply particle based kinetic simulations to explore the characteristics of a low-pressure gas discharge driven by high-voltage ($\sim$ kV) pulses with alternating polarity, with a duty cycle of $\approx$ 1\% and a repetition rate of 5 kHz. The computations allow tracing the spatio-temporal development of several discharge characteristics, the potential and electric field distributions, charged particle densities and fluxes, the mean ion energy at the electrode surfaces, etc. As such discharges have important surface processing applications, e.g. in the treatment of artificial bones, we analyse the time-dependence of the flux and the mean energy of the ions reaching the electrode surfaces, which can be both conducting and dielectric. Our investigations are conducted for argon buffer gas in the 40-140 Pa pressure range, for 1-5 cm electrode gaps and voltage pulse amplitudes ranging between 600\,V and 1200\,V.
\end{abstract}

\submitto{\PSST}

\maketitle

\section{Introduction}

\label{sec:intro}

Low-pressure gas discharges have widely been used for surface processing / modification \cite{Graves,LL,Coburn,Chen,Rath,Makabe,Donnelly,Chu,Gomathi,Fridman,Oehrlein_18PSST}. These plasma sources may operate with various types of excitation. The simplest case is represented by the direct-current (DC) excitation, which can create plasmas between conducting electrodes. Alternating-current (AC) excitation offers the possibility to create plasmas between dielectric surfaces as well. High-frequency plasma sources (e.g. capacitively and inductively coupled), which typically operate within the 1--100 MHz "radio-frequency" (RF) domain, are most widespread in, e.g. microelectronics, solar cell production, surface modification of medical implants and other areas in which plasma etching or plasma enhanced chemical vapor deposition proved to be useful~\cite{Oehrlein_18PSST}. The importance of ion energy was clearly demonstrated in several applications including plasma etching~\cite{Graves,Chen,Donnelly,Ohta01JVST}, deposition of polycrystalline silicon~\cite{Rath} and diamond like carbon films~\cite{Murakami10PRE,Michelmore15}. Although the bombardment of energetic ions is usually avoided in plasma polymerization and the role of ions is disregarded, some authors emphasize the role of ions in the mass deposition rate~\cite{Michelmore15} or cross-linking caused by the energy flux per deposited atom~\cite{Hegemann16}. Recent advancements in plasma polymerization emphasized the importance of ions for creation of radical-functionalized plasma polymers~\cite{Akhavan19}.


In RF plasmas, Voltage Waveform Tailoring \cite{Heil,Schuelgel,Lafleur,Delattre,new1}, i.e., the synthesis of complex waveforms from a harmonic signal of a given base frequency and its harmonics with given amplitudes and phases has proven to be an efficient approach for controlling the charged particle dynamics, power absorption, the generation of active species, as well as for optimising the charged particle fluxes and energy distribution functions at plasma facing surfaces both at low \cite{Derzsi,Zhang,Coumou,Diomede,Brandt} and high \cite{Gibson,Korolov} pressures. 

The time-dependence of the excitation provides an extra degree of freedom for the control of the charged particle dynamics. Both the DC and AC excitation can be continuous, modulated, or pulsed. Designing a specific waveform for the excitation may have a number of advantages. In the field of magnetron sputtering, e.g., it was realised that pulsed excitation, "High Power Impulse Magnetron Sputtering", allows creating significantly higher particle densities and fluxes as compared to DC sputtering \cite{Lundin,Anders}. In the field of gas lasers, pulsed excitation is common for cases of self-terminating transitions \cite{Little}. In capillary vacuum-ultraviolet lasers very short high-voltage pulses provide conditions for population inversion \cite{Rocca,Kukhlevsky}. Discharges with dielectric electrode surfaces ("Dielectric Barrier Discharges") exhibit self-terminating characteristics due to the charging of the insulating surfaces. Thus, despite their excitation is continuous (a simple sine wave voltage waveform) the current through the discharge flows during limited time \cite{dbd1,dbd2,dbd3}. Discharges generated by various short-pulse waveforms \cite{Pai2009,Ibuka2007,Iza2009,Lo11JPD,Ito11PRL,Huang2015,Roettgen2016,Machala,Uwe,Song,new2,new3} have found applications in combustion  \cite{combustion1,combustion2}, aerodynamic flow control \cite{flow1,flow2}, switching \cite{Bokhan,Bokhan2}, spectroscopy \cite{spec1,spec2,new8,new9}, synthesis of nanomaterials \cite{nano0}, conversion of gases \cite{Scapinello,Lotfalipour} as well as in biomedical applications through the generation of reactive species at ambient pressure and temperature \cite{Beebe,bio0,Ikawa10PPaP, Miura14JPD,Ito15PMED,Mirua16JPD}. 

A very important line of research has been focusing on the surface treatment of medical implants, where mechanical properties and biocompatibility of certain materials can significantly be improved \cite{M0,M0A}. Plasma immersion ion implantation of metallic osteosynsthesis plates has been reported in \cite{M1}. Plasma treatment has been applied as well for special alloys that are of interest in spinal deformity correction and for cardiovascular materials (used, e.g., for artificial heart valves) \cite{M2} and in tissue engineering \cite{M3}. Plasma spraying \cite{M4},  dielectric barrier discharges \cite{M5} and other types of non-thermal plasma processes \cite{new4} have been applied for the treatment of artificial bones. 

Pulsed discharges with controllable energy spectrum of ions bombarding the surfaces to be treated are promising tools for the above listed advanced applications. It has actually motivated our study that aims to uncover the connection between external control parameters (gas pressure, voltage pulse amplitude, electrode gap) and ion properties (flux, mean energy, and energy distribution) at the surfaces. To minimise the number of plasma-chemical processes, we conducted the simulations for a simple discharge gas (argon) as used in the previous plasma-processing experiments \cite{Murakami01,Kiuchi01}. However, the results bring essential implications also to future perspective plasma-chemical technologies \cite{Anjar19,Michlicek20}.

\begin{figure}[ht!]
\begin{center}
\includegraphics[width=0.96\textwidth]{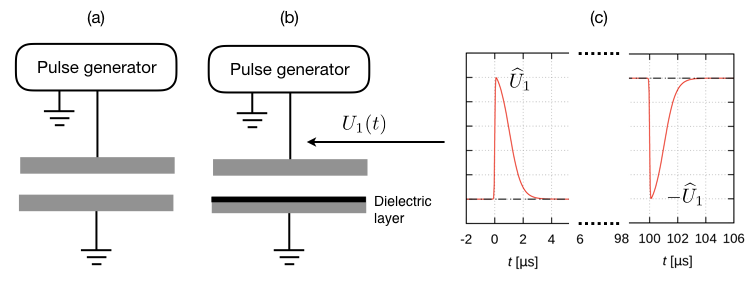}
\caption{The discharge between the parallel plate electrodes is created by a repetitive voltage pulses having alternating polarity. The upper (powered) electrode is conducting, the lower (grounded) electrode is either conducting (a) or may have a thin dielectric layer over its surface (b). 
Panel (c) shows a pair of positive/negative excitation voltage pulses, which have $\widehat{U}_1$ and $-\widehat{U}_1$ peak amplitudes, respectively, and are displaced by a delay of 100 $\mu$s. Pairs of such pulses follow each other with $f$ = 5 kHz repetition rate (i.e. the "fundamental period" is 200 $\mu$s).}
\label{fig:fig1}
\end{center}
\end{figure}

In this work, we present a numerical study of discharges established by $\sim 1\mu$s high voltage (600\,V$-$1200\,V) pulses in argon gas at pressures between 40 Pa and 140 Pa. The plasma is created between two parallel plate electrodes (see figure~\ref{fig:fig1}). The powered electrode is conducting, while the surface of the grounded electrode can be both, conducting or dielectric.  
The excitation voltage pulses follow each other with alternating polarity and 100 $\mu$s delay between the pulses. The voltage pulse shape is adopted from a measurement using the apparatus described in \cite{Sugimoto01,Takechi01}.

The paper is structured as follows. The simulation method is described in section \ref{sec:method}. The results are presented in section \ref{sec:res}, in two parts. A detailed analysis of the physics of the discharge is presented for a base set of conditions, in section \ref{sec:res1}, while the effect of the excitation voltage pulse amplitude, the gas pressure, and the electrode gap length on selected discharge characteristics are discussed in section \ref{sec:res2}. Finally, a brief summary is given in section \ref{sec:summary}. 

\section{Simulation method}

\label{sec:method}

The discharge is described by the "standard" 1d3v Particle-in-Cell simulation approach combined with the Monte Carlo treatment of collisions \cite{PIC1,PIC2,PIC3}. The working gas is argon, which is supposed to have uniform spatial density and a temperature of $T_{\rm g}$ = 350 K. The "active" species in the simulation are electrons and Ar$^+$ ions. The electron impact cross sections are adopted from \cite{Phelps1}. This set includes the elastic momentum transfer cross section, one excitation cross section that is the sum of all excitation cross sections, and the ionisation cross section. The e$^{-}+$Ar collisions are assumed to result in isotropic scattering. In the case of Ar$^+$+Ar collisions only elastic collisions are considered. The cross section set includes an isotropic scattering part, as well as a backward scattering part \cite{Phelps2}.

The electrodes of the discharge are parallel plates, situated at a distance $L$ from each other. As already mentioned in section \ref{sec:intro}, two configurations are considered: in the first case both electrodes are conducting, while in the second case one of the electrodes is supposed to be covered by a dielectric layer. This layer is modelled as a capacitor that has a value of $C$ = 10 pF for a unit area of $A_0$ = 1 cm$^2$. For both electrodes a constant ion-induced secondary electron emission coefficient ($\gamma$) is defined. For a conducting surface $\gamma$ is taken to be 0.07, while for a dielectric surface $\gamma=0.3$  is assumed. Electrons reaching the electrode surfaces are elastically reflected with a probability of $\eta$ = 0.2 \cite{Kollath}. 

The simulation uses a spatial grid with 1200 points and the length of the $T$ = 200$\,\mu$s fundamental period comprises $2 \times 10^7$ time steps.  The number of simulation particles is between 10$^5$ and $3 \times 10^5$. Convergence is typically reached during tens of periods. Due to the very low duty cycle ($\sim$ 1\%) the simulation (convergence + data collection phases) of a single case with the above settings takes several months on a single CPU. The main results of the simulations are spatiotemporal distributions of selected discharge characteristics (e.g. the electron density and the ionisation source function), as well as the flux and the energy distribution of the ions hitting the electrode surfaces. These characteristics are studied as a function of the operating conditions (nature of electrode surfaces, gas pressure, electrode gap, and voltage peak amplitude). 

\section{Results} 

\label{sec:res}

In the following section \ref{sec:res1}, we illustrate the main discharge characteristics for the case of $\widehat{U}_1$ = 1000\,V, $p$ = 100\,Pa, and $L$ = 3\,cm, which is considered here as the "base case". Subsequently in section \ref{sec:res2}, we present a parameter variation, where we study the effects of the gas pressure, the voltage pulse amplitude, as well as the electrode separation.

\subsection{Characterisation of the discharge behaviour}

\label{sec:res1}

Here, for the "base case" defined above, we compare the properties of the  discharge with different electrodes. The powered electrode is taken to be conductive in all cases, and is characterised with a secondary electron emission coefficient of $\gamma_{\rm p}=0.07$ \cite{PP99,gamma_CAST}. This value is typical for Ar$^+$ ion bombardment of metal surfaces at moderate energies up to hundreds of eV-s \cite{PP99}. For the other electrode that is connected to ground potential, we consider both conducting and dielectric surfaces. In the former case, we assume $\gamma_{\rm g}= \gamma_{\rm p} =0.07$, while in the latter we chose a higher secondary electron emission yield typical for dielectric materials, $\gamma_{\rm g} =0.3$. (The "p" and "g" subscripts refer to the powered and grounded electrodes, respectively.) 

For an additional test case, $\gamma_{\rm g} =0.07$ is used instead of $\gamma_{\rm g} =0.3$ for the grounded electrode with a dielectric layer. This test is supposed to check the correctness of the simulation: at equal secondary yields at both electrodes it is irrelevant which surface is covered by the dielectric layer. Thus time-shifted but spatially symmetrical patterns are expected from the simulation.

\begin{figure}[ht!]
\begin{center}
\includegraphics[width=0.48\textwidth]{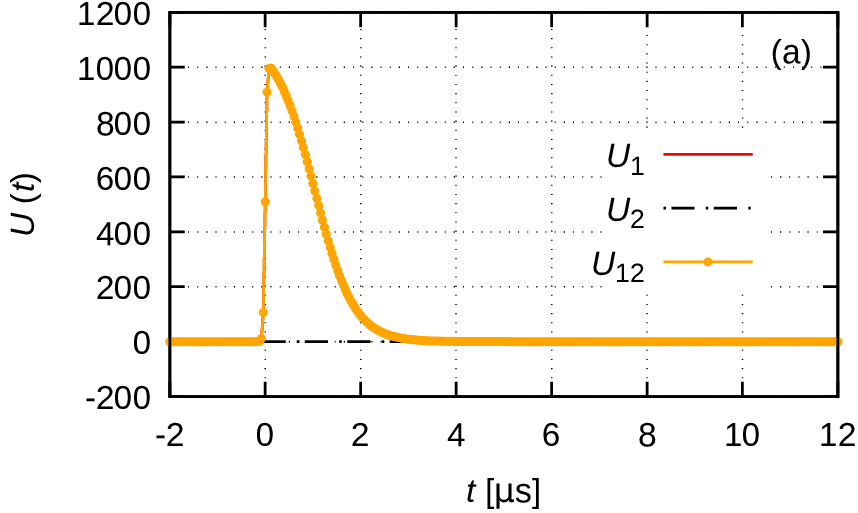}~~~~
\includegraphics[width=0.48\textwidth]{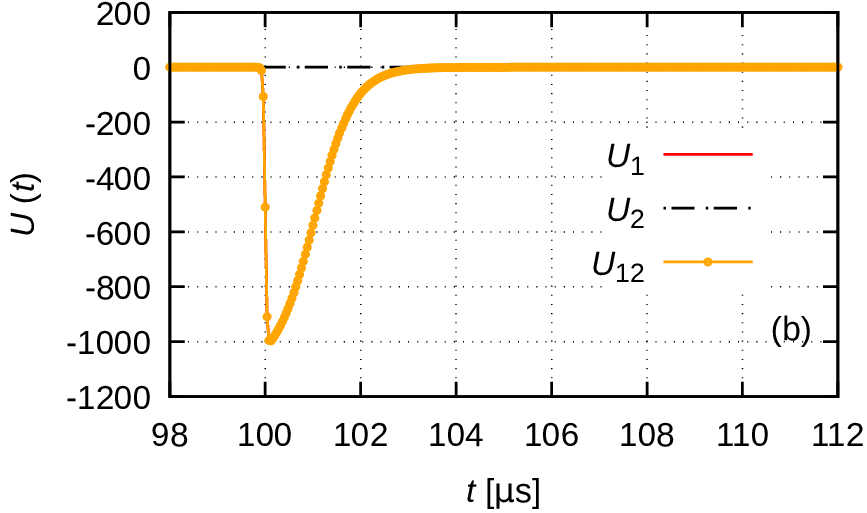}\\
\includegraphics[width=0.48\textwidth]{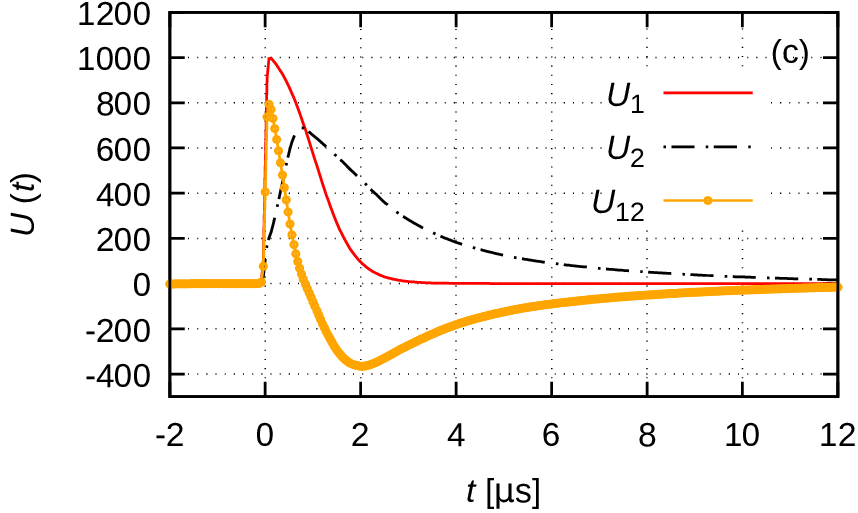}~~~~
\includegraphics[width=0.48\textwidth]{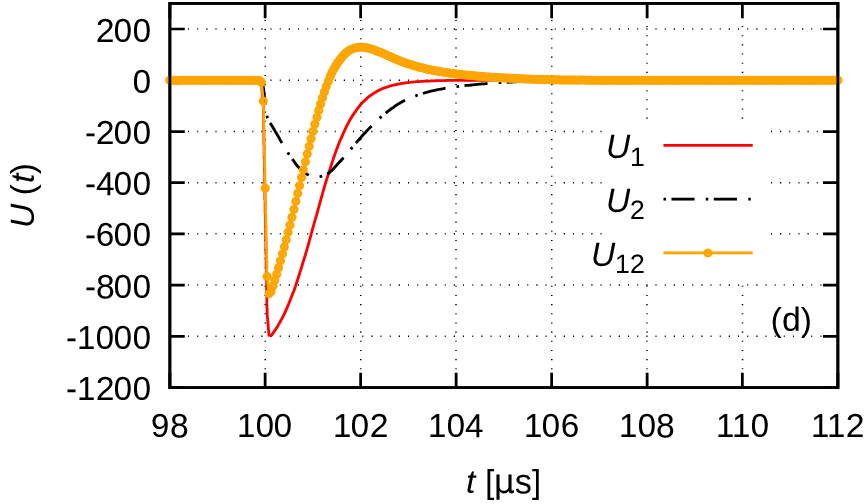}\\
\includegraphics[width=0.48\textwidth]{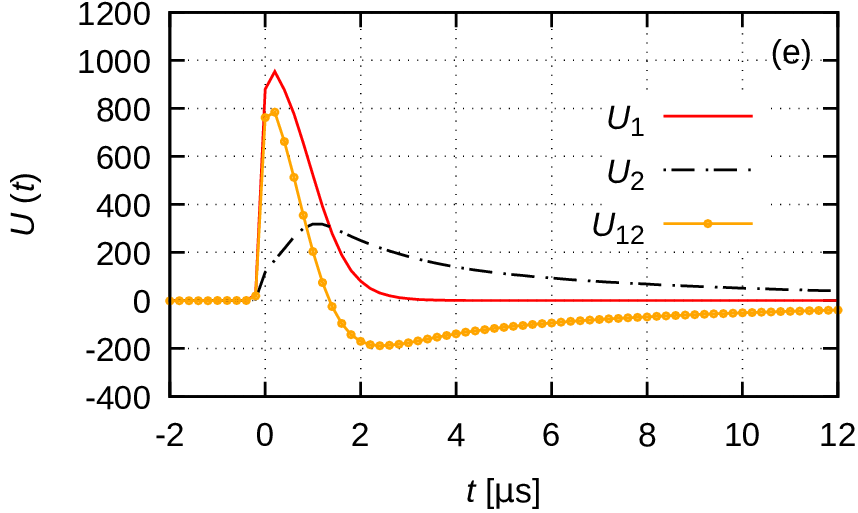}~~~~
\includegraphics[width=0.48\textwidth]{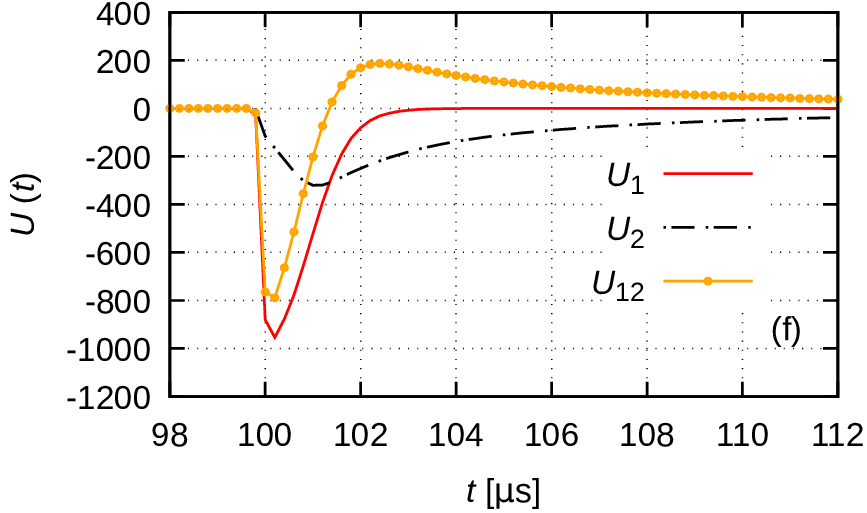}\\
\caption{Time dependence of the potentials of the electrodes. $U_1$ equals the applied voltage. In the case of a conducting grounded electrode $U_2$ = 0. If there is a dielectric layer on the grounded electrode, $U_2$ is the surface potential. $U_{12} = U_1 - U_{2}$ is the discharge voltage. The potentials are shown for (a,b) both electrodes conducting, (c,d) conducting powered electrode and dielectric layer on the grounded electrode, with $\gamma_{\rm p}=0.07$ and $\gamma_{\rm g}=0.3$, (e,f) the same as (c,d) with $\gamma_{\rm p}=0.07$ and $\gamma_{\rm g}=0.07$. Panels in the left column correspond to the positive polarity applied voltage pulses, while panels in the right column correspond to negative pulses. The peak value of the excitation is $\widehat{U}_1 = \pm$ 1000\,V, $p$ = 100\,Pa, and $L$ = 3\,cm.}
\label{fig:voltages}
\end{center}
\end{figure}

Figure~\ref{fig:voltages} shows the relevant voltage waveforms at the electrode surfaces and over the discharge gap. Panels in the left column (a,c,e) present the data for the vicinity of the positive excitation pulse, while panels in the right column (b,d,f) shows the same data in the vicinity of the negative excitation pulse. The voltage waveform applied to the powered electrode, $U_1$, has in this "base" case $\pm$1000 V peak amplitude and a width of $\approx$\,1\,$\mu$s. The shape of the excitation waveform has been adopted from experiments \cite{Sugimoto01}, and is kept the same throughout our studies, only the peak amplitude is varied in section \ref{sec:res2}. The positive and negative pulses have exactly the same shape, only their polarity differs. The positive pulse is applied at $t=0 \, \mu$s, whereas the negative pulse appears at $t=100\, \mu$s. The pulses repeat with the (fundamental) period of 200 $\mu$s, corresponding to a repetition rate of $f$ = 5 kHz that is typically used in the current experiments \cite{Anjar19,Michlicek20}. 

Figures~\ref{fig:voltages}(a) and (b) show the case of a conducting grounded electrode. The potential of this electrode, $U_2$, is zero, and the voltage $U_{12}$ over the discharge equals $U_1$, the potential applied to the powered electrode. Figures~\ref{fig:voltages}(c) and (d) depict the positive and negative pulses, respectively, for a grounded electrode with a dielectric surface. These results were obtained with $\gamma_{\rm p}=0.07$ and $\gamma_{\rm g}=0.3$. Following the application of the high voltage pulses, the surface of the grounded electrode charges up in a way that the discharge voltage becomes lower in magnitude than the applied voltage. Beyond $\approx 1.4\, \mu$s after the peaks of the excitation pulses $|U_2|$ becomes higher than $|U_1|$, and consequently, the discharge voltage, $U_{12}$, changes sign. In the case of the positive excitation pulse, this "reversed voltage" has a magnitude of about $-$400 V, while for the negative pulse, it amounts to about 150 V. This asymmetry originates from the different values of the secondary electron yields at the two electrodes. The symmetry of the discharge is re-established when equal $\gamma$ values are used, as figures~\ref{fig:voltages}(e) and (f) show for a pair of conducting/dielectric electrodes with $\gamma_{\rm p}=\gamma_{\rm g}=0.07$. Of course, the value for the dielectric surface is unrealistically low. It was assumed only to confirm the correctness of the simulations, which are supposed to give symmetric results because the dielectrics is modelled as a capacitor that can reside anywhere in the electrical circuit. 

\begin{figure}[ht!]
\begin{center}
\includegraphics[width=0.5\textwidth]{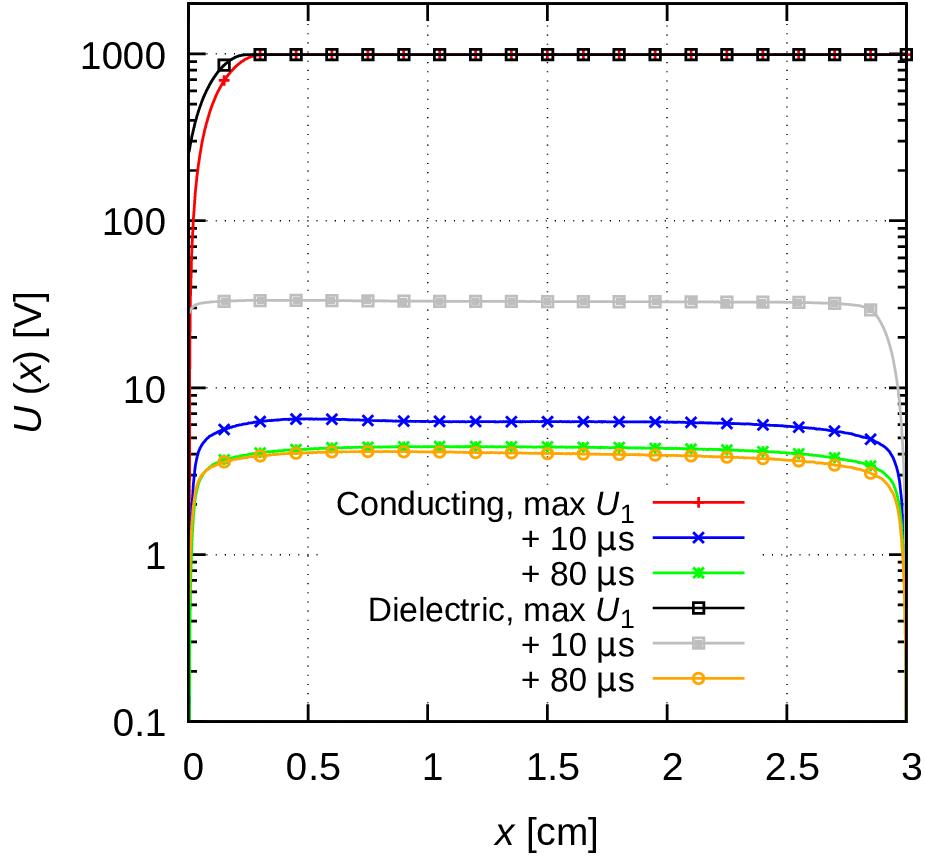}
\caption{Potential distribution in the discharge for conducting and dielectric grounded electrodes, in the case of a positive excitation pulse. "max $U_1$" is the potential at the time of the maximum applied voltage, other curves show the potential distributions at later times specified in the legend. "Base case" discharge conditions: $\widehat{U}_1$ = 1000\,V, $p$ = 100\,Pa. The grounded electrode is located at $x$ = 0\,cm, while the powered electrode is at $x=L=$ 3\,cm.}
\label{fig:pot_distr}
\end{center}
\end{figure}

Figure \ref{fig:pot_distr} shows the potential distribution in the discharge for both the cases of the conducting and dielectric grounded electrodes following the application of a positive excitation pulse. The curves marked as "max $U_1$" correspond to the time of the maximum applied voltage, and the other curves show the potential distributions  10 and 80\,$\mu$s later. At the maximum applied voltage we observe the sheath formation at the grounded electrode ($x$ = 0\,cm). At this time, the plasma potential is a few Volts higher (not visible in the figure) compared to the potential of the powered electrode. At later times, the plasma potential decreases. This decrease is faster in the case of the conducting electrode, where after 10\,$\mu$s we find a value of about 5\,V, whereas the plasma potential is $\sim$ 30\,V in the case of the dielectric electrode. Note, however, that in the dielectric case the sheath "moves" to the powered electrode at this time. Even later, the potential drops to about 3\,V, for both types of the electrodes. The resulting ambipolar electric field and potential drops near both electrodes retain the electrons in the (afterglow) plasma between the pulses, as well as accelerate the ions towards the electrodes to compensate for the flux of fast electrons escaping from the plasma. The electric field near the electrode surfaces is in the order of $\sim$ 80\,V/cm during the late afterglow.   

\begin{figure}[ht!]
\begin{center}
\includegraphics[width=0.45\textwidth]{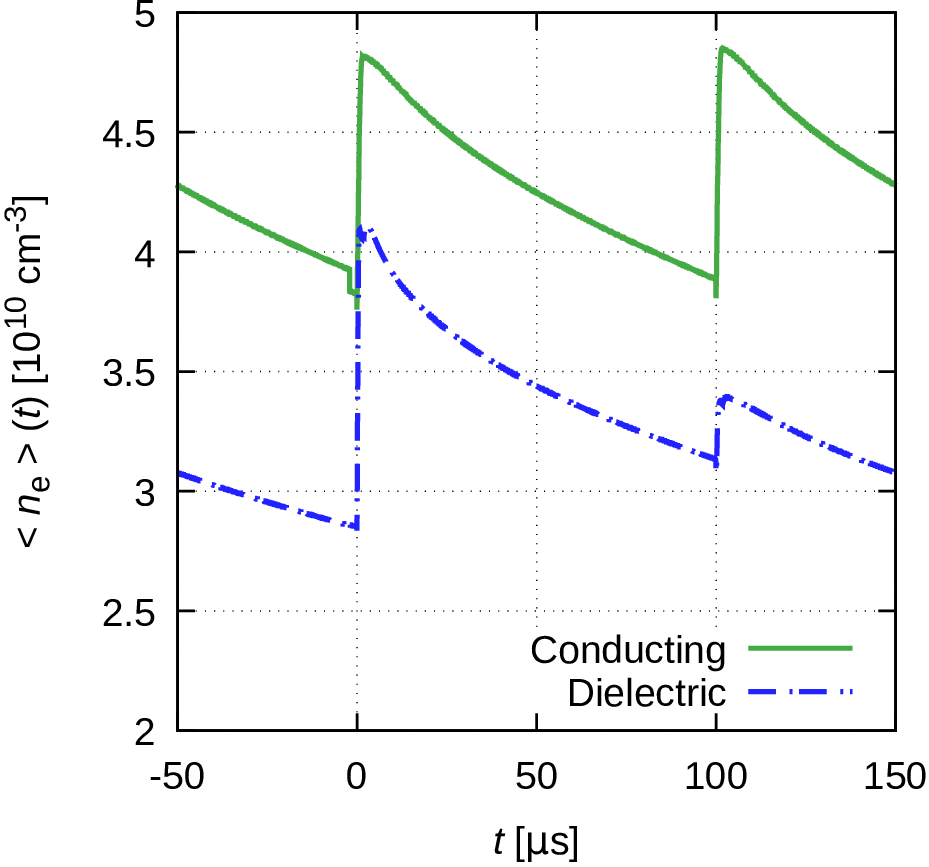}
\caption{Time dependence of the spatially averaged electron density in the cases of conducting and  dielectric grounded electrode. "Base case" discharge conditions: $\widehat{U}_1$ = 1000\,V, $p$ = 100\,Pa, and $L$ = 3\,cm.}
\label{fig:av-densities}
\end{center}
\end{figure}

The markedly different behaviour of the gap voltage in the cases of a conducting vs. a dielectric grounded electrode, revealed in figure~\ref{fig:voltages}, has consequences on the charged particle dynamics and the discharge characteristics. For example, the temporal evolution of the mean (i.e. spatially averaged) electron density is significantly different in these two cases, as it can be seen in figure~\ref{fig:av-densities}.

In the case of a conducting grounded electrode the average electron density varies between $3.8 \times 10^{10}$ cm$^{-3}$ and $4.8 \times 10^{10}$ cm$^{-3}$, for the base conditions of $\widehat{U}_1$ = 1000\,V, $p$ = 100\,Pa, and $L$ = 3\,cm. Upon the appearance of the excitation pulses a very sharp increase is observed, which is followed by a slow decay. The dynamics is the same for the positive and negative excitation pulses. Between the high-voltage excitation pulses the electron density is retained by the ambipolar electric field that builds up in the gap. At the relatively high pressure, a large amount of the charged particles survives in the gap between the excitation pulses.  

\begin{figure}[ht!]
\begin{center}
\includegraphics[width=0.48\textwidth]{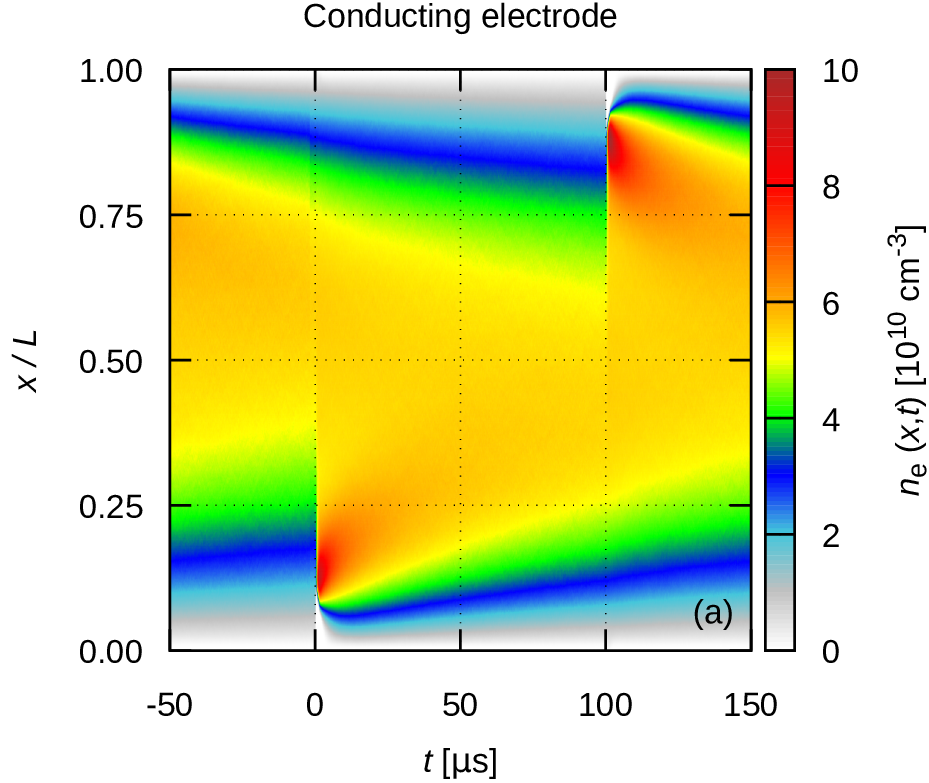}~~~~
\includegraphics[width=0.48\textwidth]{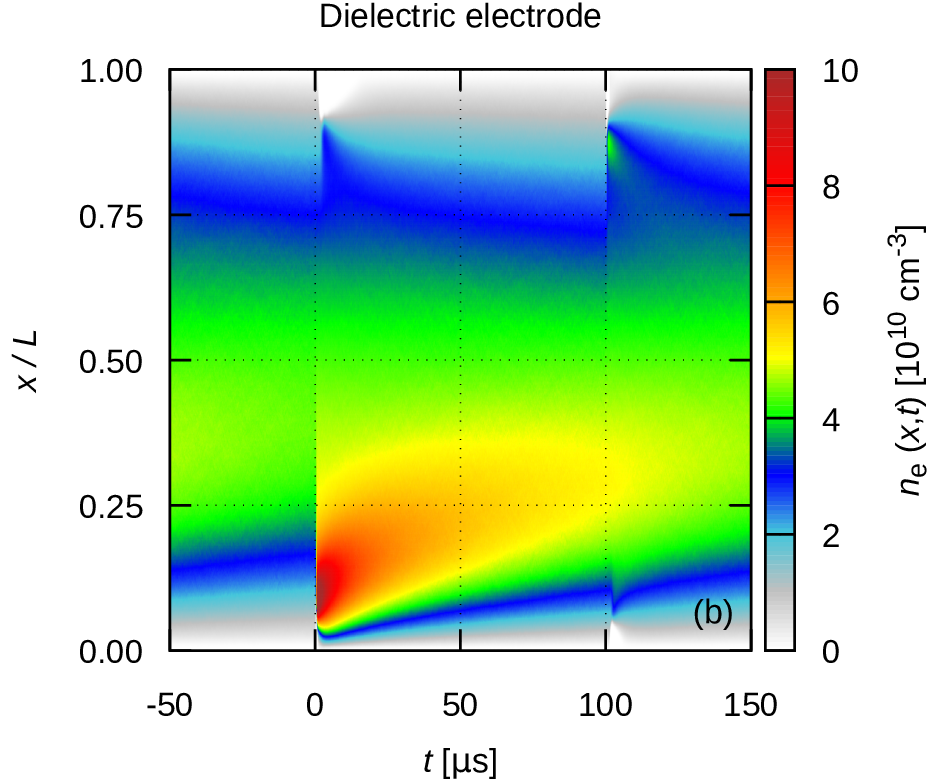}
\caption{Spatio-temporal distribution of the electron density in the cases of (a) conducting and (b) dielectric grounded electrode. "Base case" discharge conditions: $\widehat{U}_1$ = 1000\,V, $p$ = 100\,Pa, and $L$ = 3\,cm. The grounded electrode is located at $x/L=0$, while the powered electrode is at $x/L=1$.}
\label{fig:st-densities}
\end{center}
\end{figure}

For the dielectric grounded electrode, on the other hand, very different behaviour is found for the pulses with opposite polarity, which is a consequence of the different values of the secondary electron yields at both electrodes. The positive pulse results in an approximately four times higher "jump" of the density as compared to that found for the negative pulse. It can be explained by the fact that upon the application of the positive pulse, the grounded electrode with the dielectric layer acts as the temporary cathode, where the charge reproduction mechanisms are largely enhanced due to the higher $\gamma$. In this case, the average electron density varies between $2.8 \times 10^{10}$ cm$^{-3}$ and $4.15 \times 10^{10}$ cm$^{-3}$. Further information about the spatio-temporal behaviour of the electron density is revealed in figure~\ref{fig:st-densities}.

In the case of conducting grounded electrode (see  figure~\ref{fig:st-densities}(a)) symmetrical patterns (mirrored in space and shifted in time) of the spatial distribution of the electron density can be observed. Upon the appearance of the excitation pulses a significant increase of the electron density occurs. The positive excitation pulse results in an electron density increase near the grounded electrode (situated at $x/L=0$), while the negative excitation pulse results in the buildup of high electron density near the powered electrode (situated at $x/L=1$), which act as the temporary cathode at the given times and voltage polarities. Following the termination of the excitation pulses the electron density spreads towards the centre of the discharge. In the case of the dielectric grounded electrode a remarkable asymmetry is observed between the negative and positive excitation pulses, as it can be seen in figure~\ref{fig:st-densities}(b). 

\begin{figure}[ht!]
\begin{center}
\includegraphics[width=0.46\textwidth]{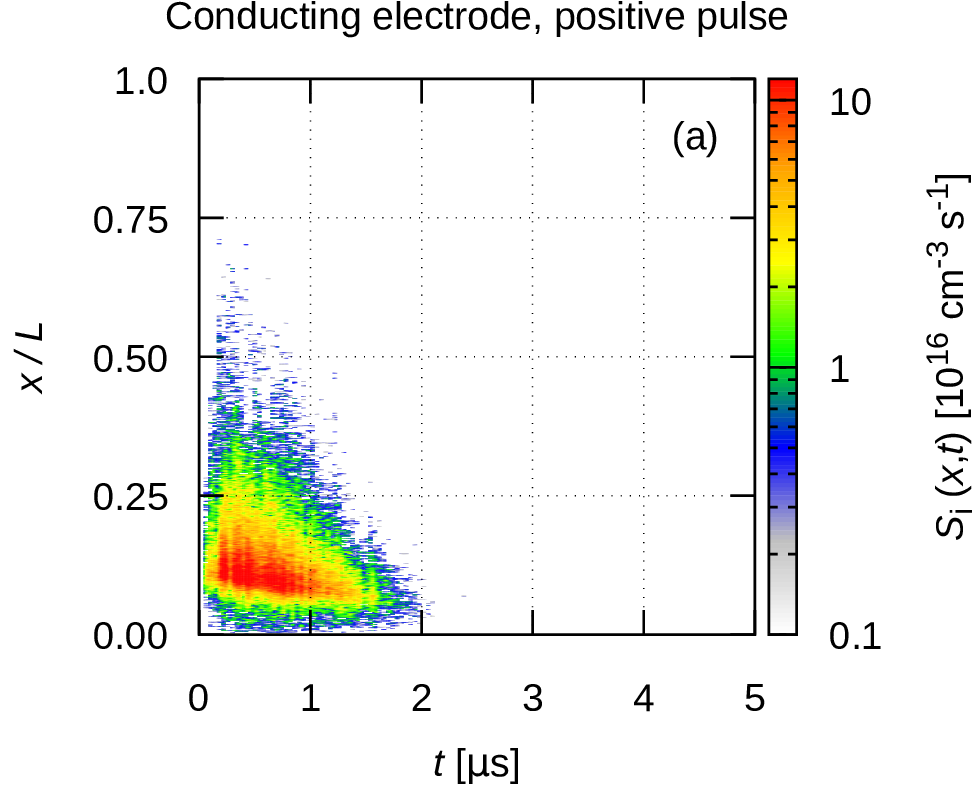}~~~~
\includegraphics[width=0.46\textwidth]{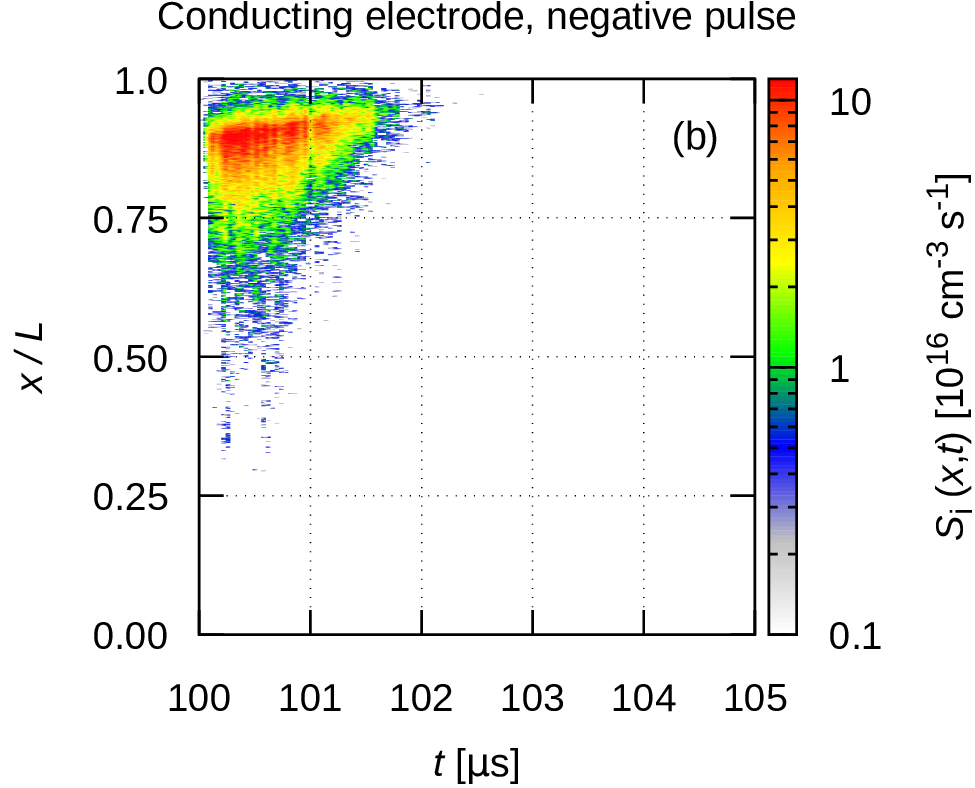}\\
\vspace{0.5cm}
\includegraphics[width=0.46\textwidth]{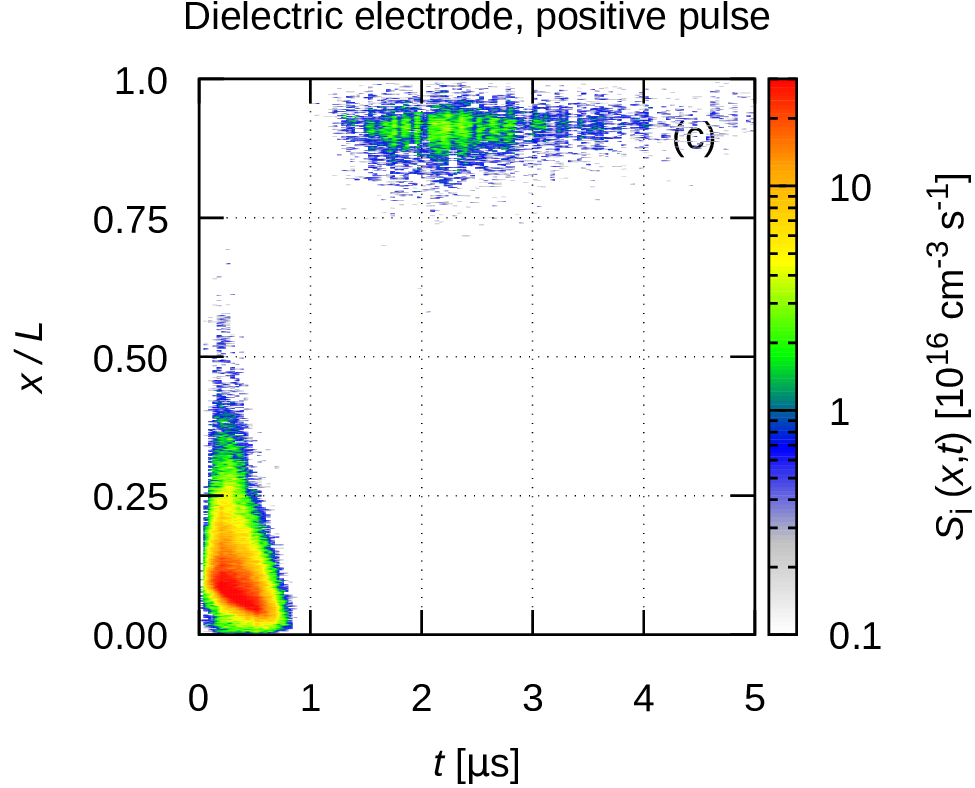}~~~~
\includegraphics[width=0.46\textwidth]{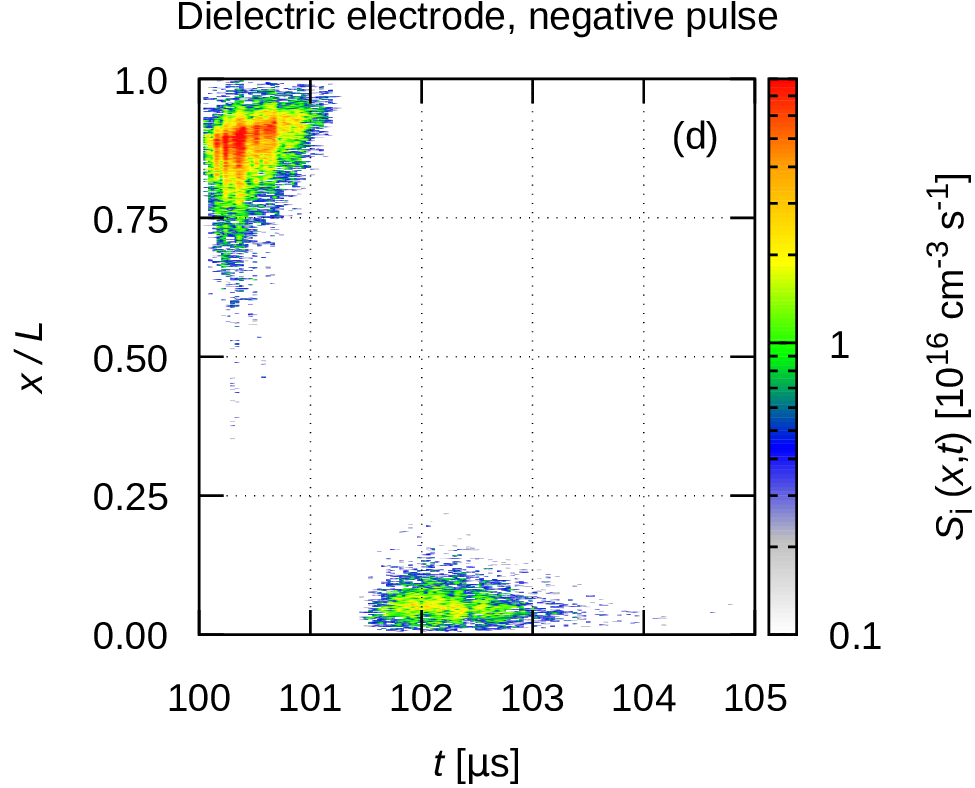}\caption{Spatio-temporal distribution of the ionisation source function, $S_{\rm i}(x,t)$, in the case of (a,b) conducting and (c,d) dielectric grounded electrode. Panels in the left column correspond to the positive polarity applied voltage pulses, while panels in the right column correspond to negative pulses. "Base case" discharge conditions: $\widehat{U}_1$ = 1000\,V, $p$ = 100\,Pa, and $L$ = 3\,cm. The grounded electrode is located at $x/L=0$, while the powered electrode is at $x/L=1$.}
\label{fig:st-ionisation}
\end{center}
\end{figure}

It is a peculiarity of the discharge with a dielectric electrode that an increase of the electron density is also observed at the anode side upon the application of the high-voltage pulses. Such an increase is not seen for the conducting grounded electrode, in figure~\ref{fig:st-densities}(a). The reason for this is the reversed potential (c.f. figures \ref{fig:voltages}(c) and (d)), the effect of which on the ionisation is studied in figure~\ref{fig:st-ionisation}. The (a) and (b) panels of this figure show the ionisation source function, $S_{\rm i}(x,t)$, for the conducting electrode, while panels (c) and (d) present the dielectric case. 

In the case of the conducting grounded electrode, significant ionisation is observed during $\approx 1.5\,\mu$s, i.\,e., close to the width of the excitation pulses. Ionisation is confined to within approximately one-fourth of the gap, in the vicinity of the temporary cathode.

For the dielectric grounded electrode, the ionisation is confined to a narrower time interval ($\lesssim 0.8\mu$s), because the charging of the dielectrics leads to the self-termination of the discharge. Due to the appearance of a reversed potential over the electrode gap as a consequence of this charging process a weaker, but still significant ionisation is also observed at the "opposite" electrode, as figures~\ref{fig:st-ionisation}(c,d) reveal. Ionisation in these domains of space and time is weaker compared to that caused by the primary pulse.

\begin{figure}[ht!]
\begin{center}
\includegraphics[width=0.42\textwidth]{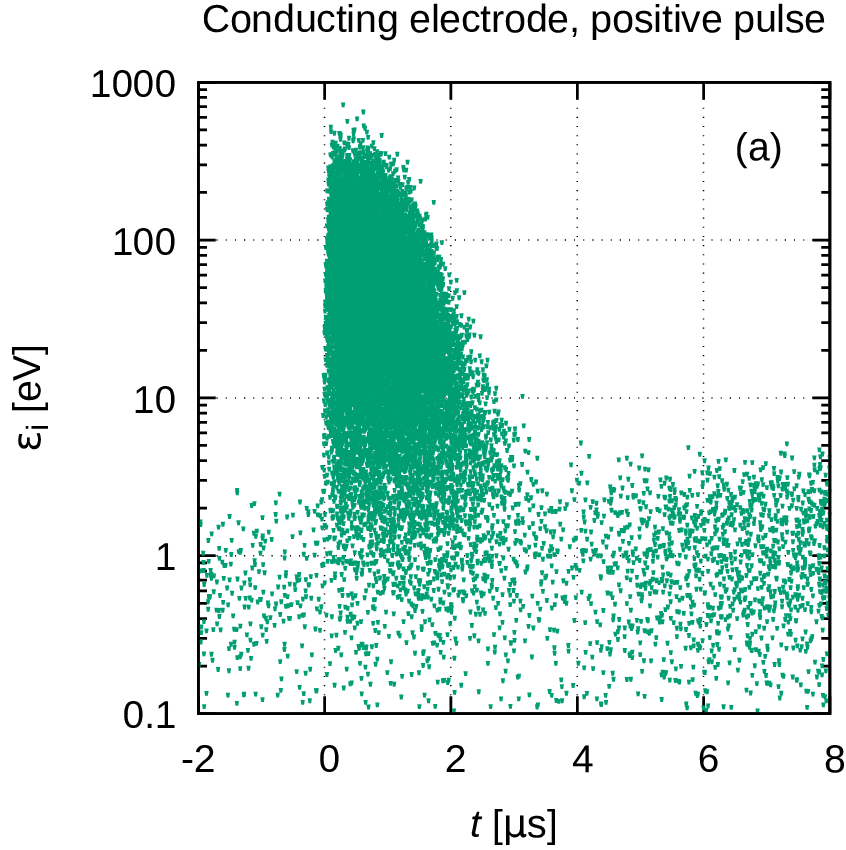}~~~
\includegraphics[width=0.43\textwidth]{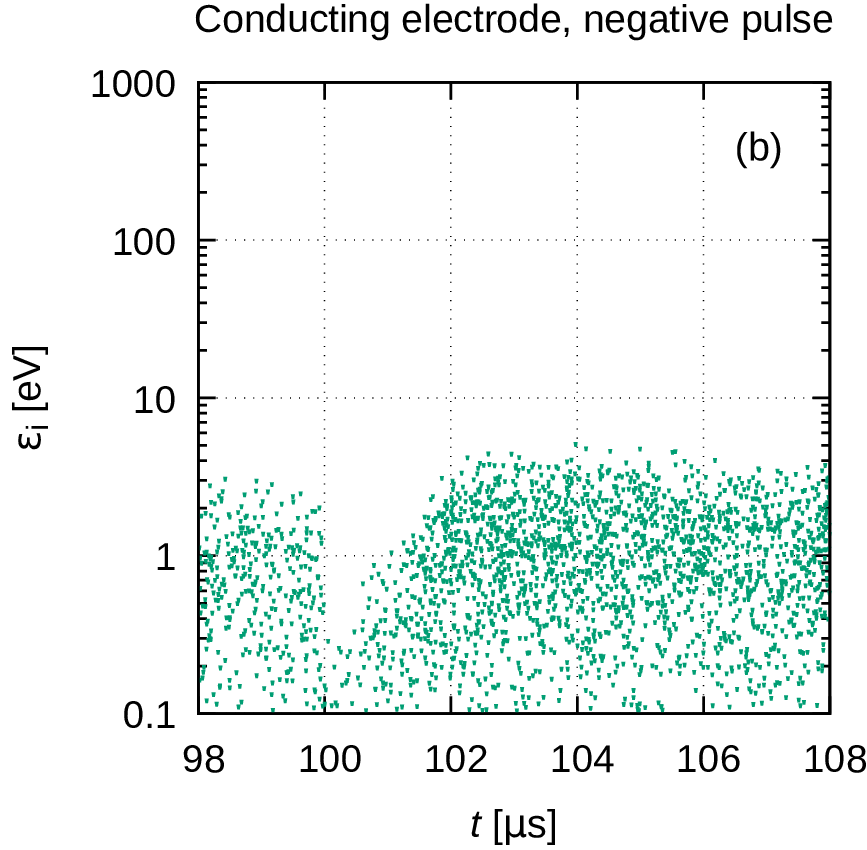}\\
\vspace{0.25cm}
\includegraphics[width=0.42\textwidth]{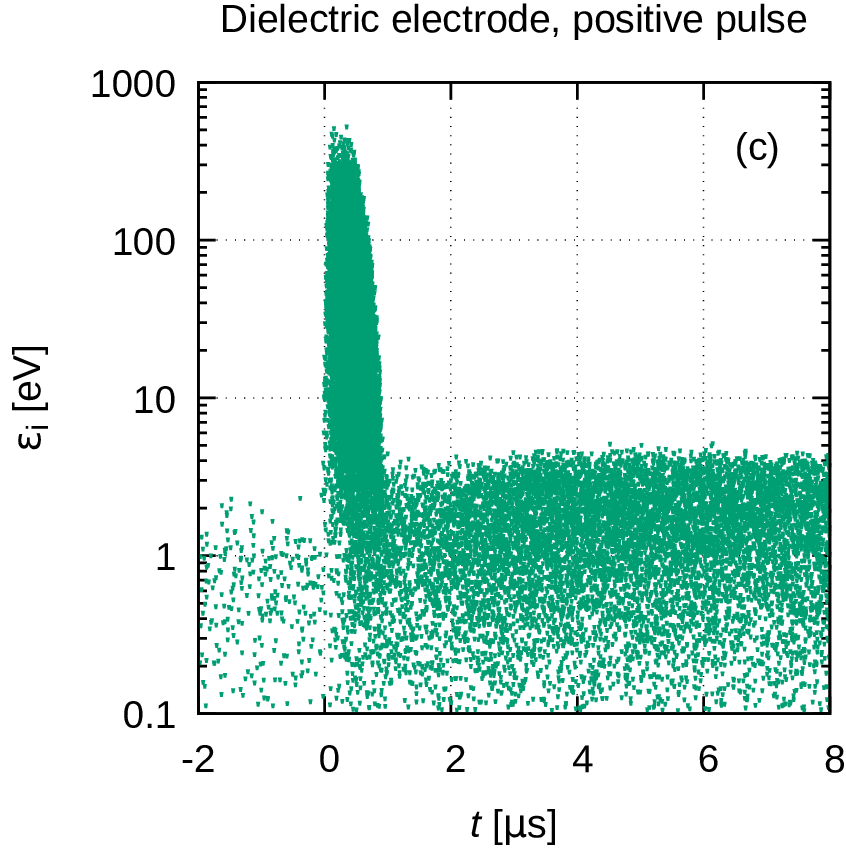}~~~
\includegraphics[width=0.43\textwidth]{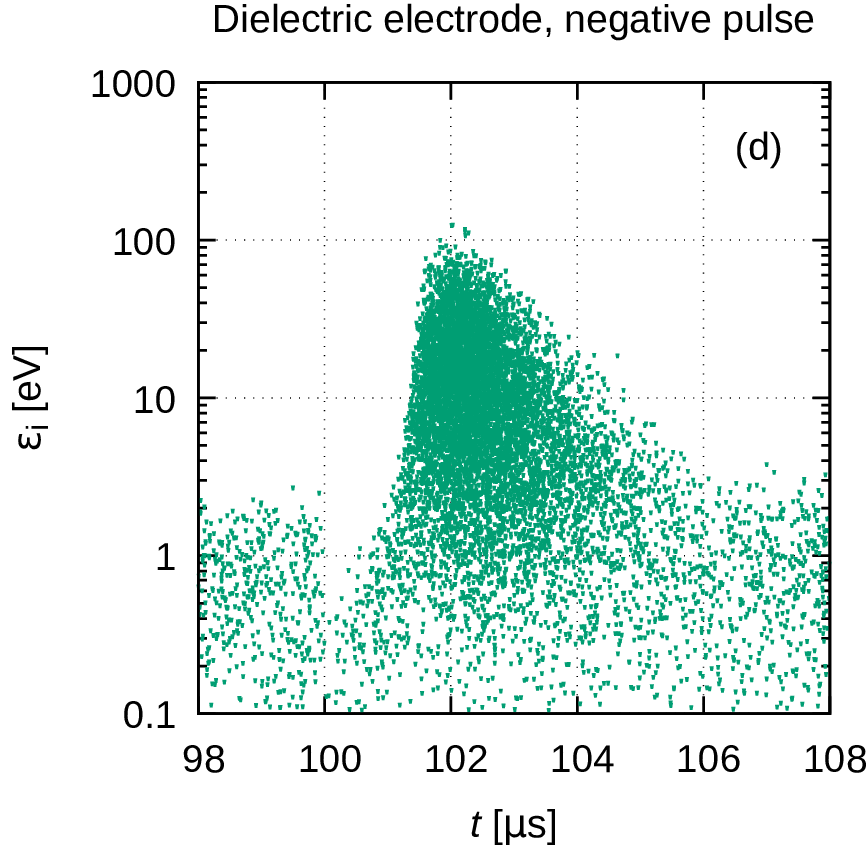}\\
\caption{Energy of individual ions (represented as single dots) reaching the grounded electrode, in the case of a conducting (a,b) and dielectric surface (c,d). Panels (a) and (c) refer to the vicinity of the positive excitation pulse, while (b) and (d) show the distributions in the vicinity of the negative excitation pulse. "Base case" discharge conditions: $\widehat{U}_1$ = 1000\,V, $p$ = 100\,Pa, and $L$ = 3\,cm.}
\label{fig:arrival}
\end{center}
\end{figure}

In plasma processing of materials, the energy distribution of the ions reaching the electrodes is very important. In the following, we examine the energy of the Ar$^+$ ions at the grounded electrode. 
In the panels of figure~\ref{fig:arrival}, each arriving ion is represented by a dot. Panels (a) and (b) in the first row present the case of the conducting grounded electrode, within a time "window" of 10 $\mu$s. Due to the rapidly increasing voltage near $t=0\,\mu$s, the energy of the ions increases promptly and ions with up to $\sim$\,500\,eV are found right after applying the positive excitation pulse. (Recall that in this case, the grounded electrode is the temporary cathode.) Streaming of the high-energy ions to the surface lasts for about 2\,$\mu$s. Subsequently, the energy of the ions falls below $\sim$ 5\,eV. Low energy ions, accelerated by the ambipolar electric field, continue to stream to the electrode  until the second excitation pulse with negative polarity arrives at $t=100\,\mu$s. Upon the arrival of the negative excitation pulse the ion energy and the number of incoming ions are slightly depleted at the grounded electrode because it acts as the temporary anode. A few $\mu$s later the low energy Ar$^+$ ion flow to the electrode is re-established.

In the case of the dielectric grounded electrode, the pulse of the high-energy ions due to the positive excitation pulse (figure~\ref{fig:arrival}(c)) lasts for a shorter time as compared to the conducting grounded electrode case. Here, the ions reach about the same energy but their flux (proportional to the density of dots in the representation used in figure~\ref{fig:arrival}) after the excitation pulse is much higher compared to the conducting case. At the time of the negative excitation pulse (see figure~\ref{fig:arrival}(d)), the ion energy and flux are first depleted similarly to the case of the conducting electrode. However, a bunch of high-energy ions is observed following the negative voltage pulse, as expected under the reversal of the gap voltage (discussed earlier).

\begin{figure}[ht!]
\begin{center}
\includegraphics[width=0.45\textwidth]{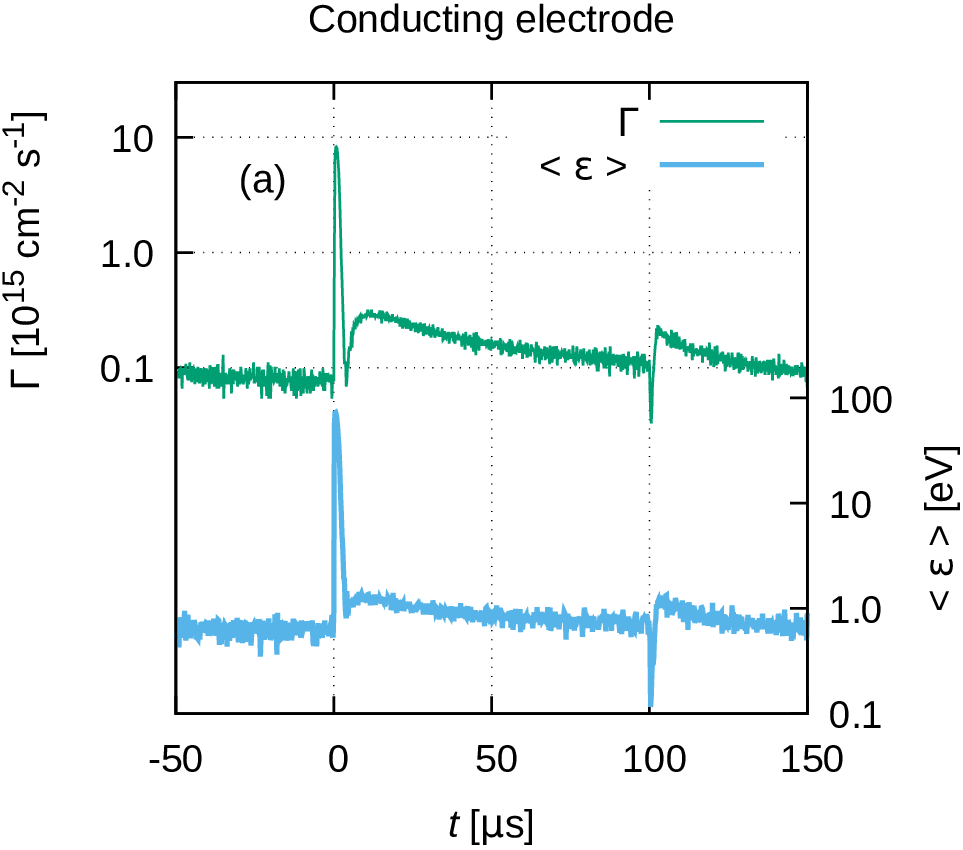}~~~~
\includegraphics[width=0.45\textwidth]{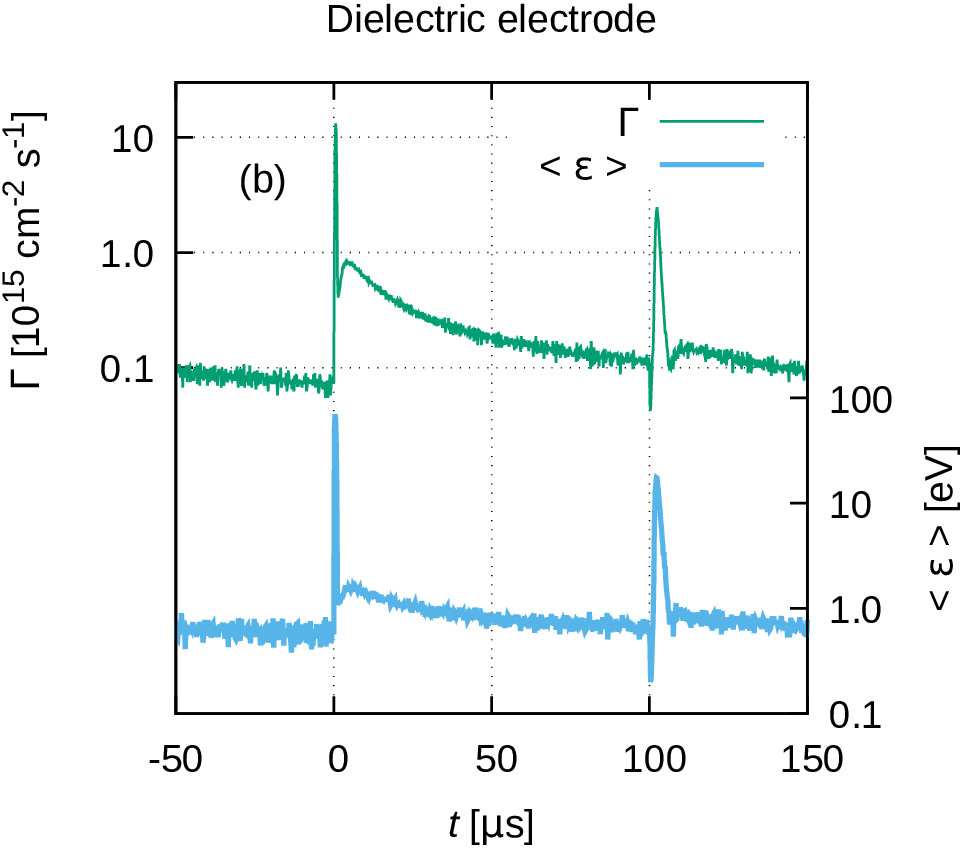}
\caption{Flux and mean energy of Ar$^+$ ions reaching the grounded electrode as a function of time. The grounded electrode had either conducting (a) or dielectric (b) surface.}
\label{fig:ionprop}
\end{center}
\end{figure}

The flux and the mean energy of Ar$^+$ ions for the conducting and  dielectric grounded electrodes are shown in figures~\ref{fig:ionprop}(a) and (b), respectively. In both cases, ion fluxes are in the order of 10$^{15}$\,cm$^{-2}$\,s$^{-1}$, and the mean ion energy approaches 100 eV upon the application of the positive excitation pulses. This value drops to $\lesssim$ 1\,eV between the excitation pulses.

The energy of the ions arriving during the afterglow periods to the electrodes is determined by the strength of the ambipolar electric field discussed in relation to figure \ref{fig:pot_distr}. The cross section for the Ar$^+$+Ar collisions at low ion energies is about $\sigma_{\rm i} \approx 5 \times 10^{-15}$ cm$^2$ \cite{Phelps2}. At the given values of the gas pressure and temperature, the gas density is $n_{\rm g} \approx 2.1 \times 10^{16}$ cm$^{-3}$. It results in a mean free path of the
ions in the order of 0.01 cm. The ambipolar electric field of $\sim$ 80\,V/cm found for the base conditions near the electrode surfaces results in the mean ion energy of $\approx$ 0.8\,eV in between the pulses, and this is clearly the value that we observe in figure \ref{fig:ionprop}. 

\subsection{Effects of the operating conditions}

\label{sec:res2}

In this section, we present simulation results for the time-dependence of the mean electron density, as well as for the flux and the mean energy of the Ar$^+$ ions streaming to the grounded electrode of the discharge, as a function of the excitation voltage pulse amplitude, the gas pressure, and the electrode gap. 

\begin{figure}[ht!]
\begin{center}
\includegraphics[width=0.36\textwidth]{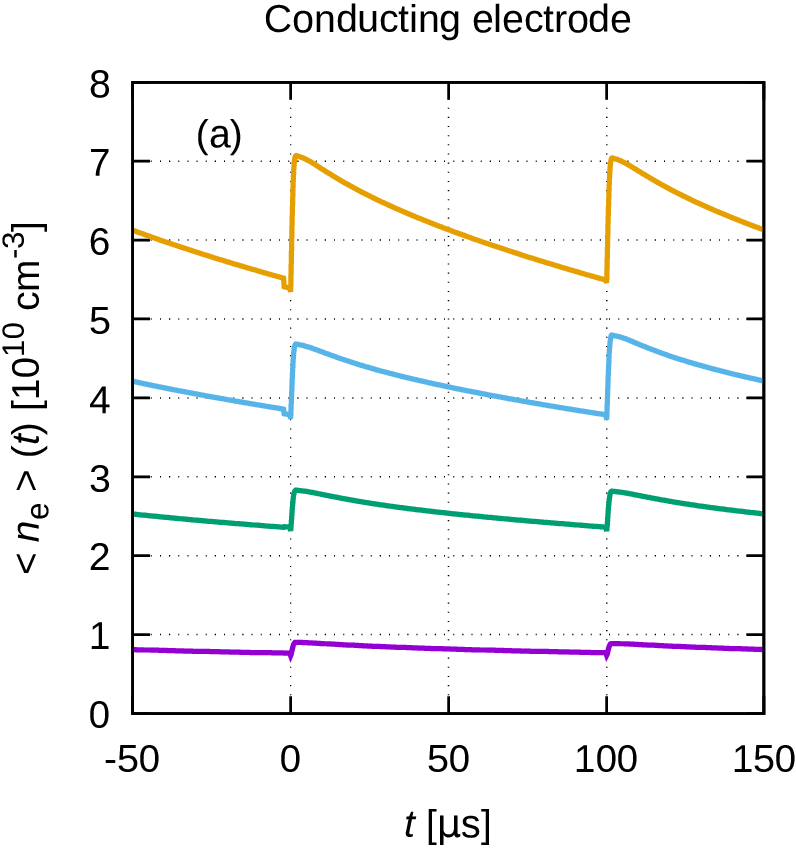}
~~~~~~~~~~~~~~~\includegraphics[width=0.36\textwidth]{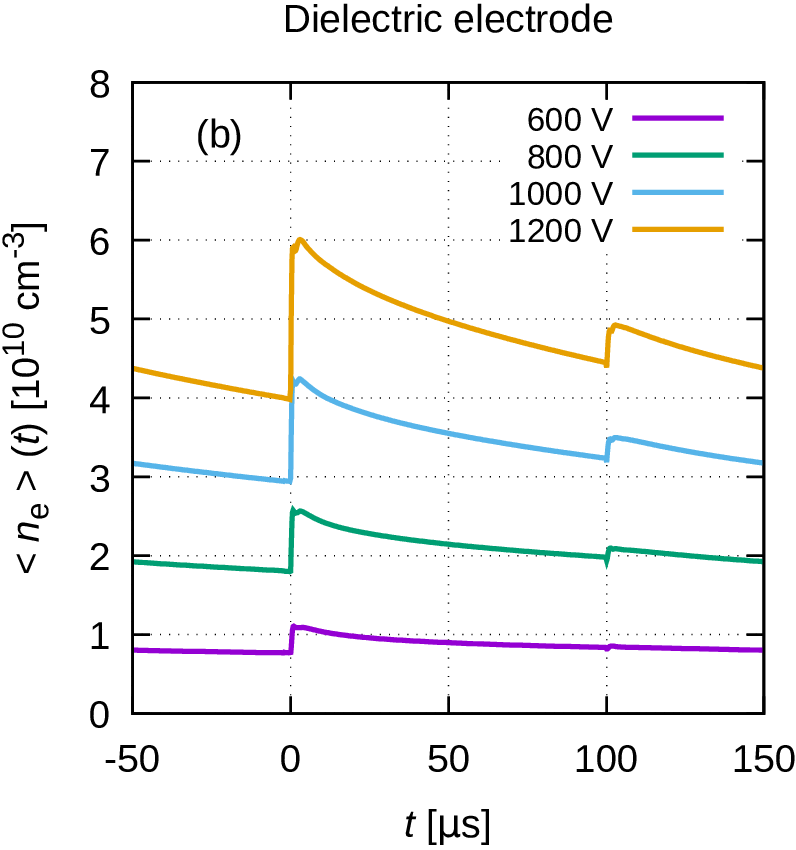}\\
\vspace{0.5cm}
~~~~\includegraphics[width=0.45\textwidth]{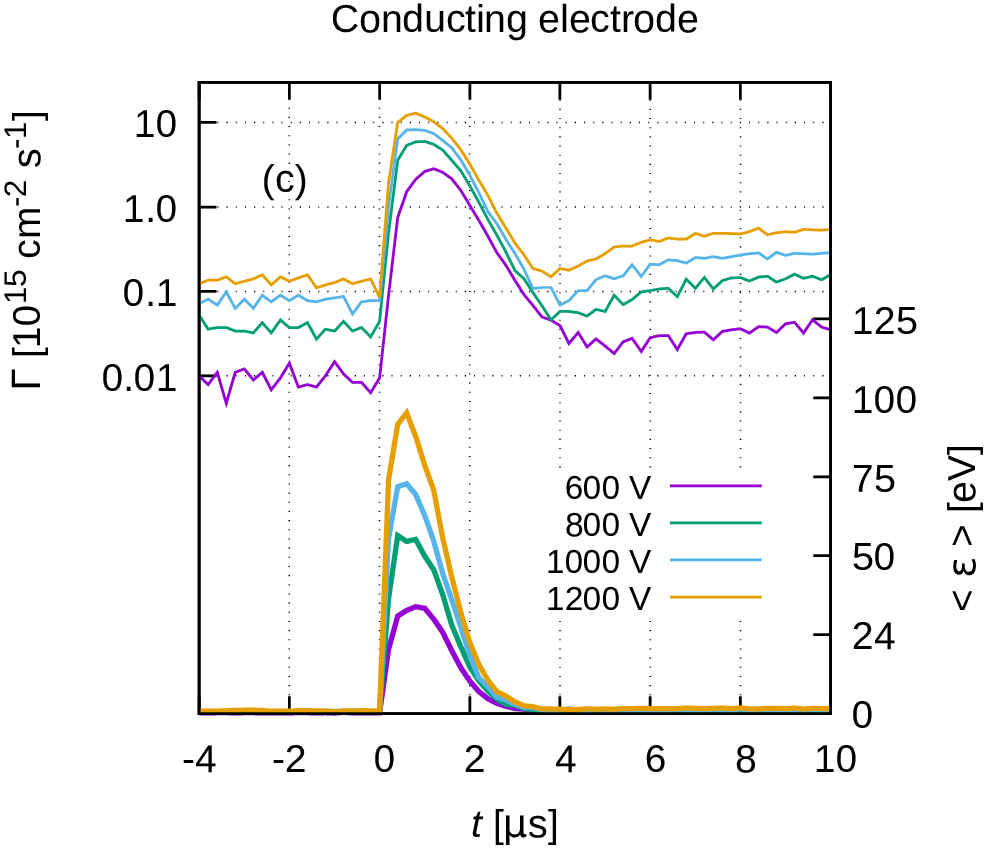}~~~~
\includegraphics[width=0.45\textwidth]{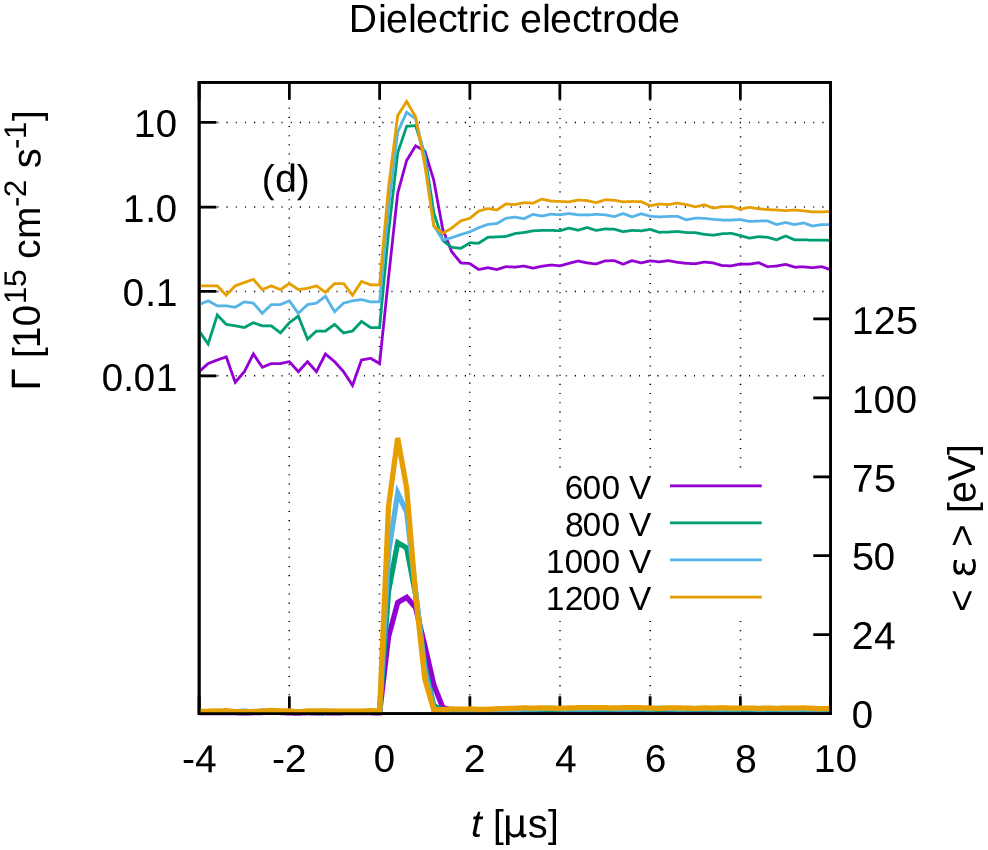}
\caption{Effect of the driving voltage pulse amplitude on the discharge characteristics: (a,b) mean electron density as a function of time; (c,d) flux (left scale, upper set of thin lines) and mean energy (right scale, lower set of thick lines) of Ar$^+$ ions reaching the grounded electrode as a function of time. Results for the case of a conducting grounded electrode are shown in the left column, while results for the case of a dielectric electrode are shown in the right column. $p$ = 100 Pa, $L$ = 3 cm, $\gamma = 0.07$ for the conducting surfaces and $\gamma = 0.3$ for the dielectric surface.}
\label{fig:par1}
\end{center}
\end{figure}

\begin{figure}[ht!]
\begin{center}
\includegraphics[width=0.36\textwidth]{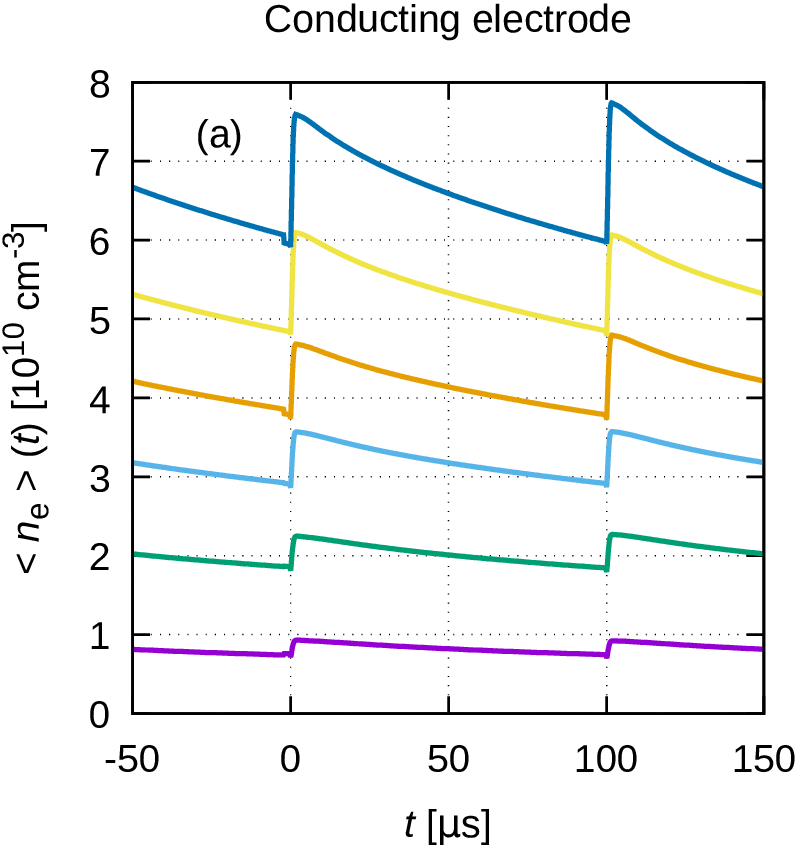}
~~~~~~~~~~~~~~~\includegraphics[width=0.36\textwidth]{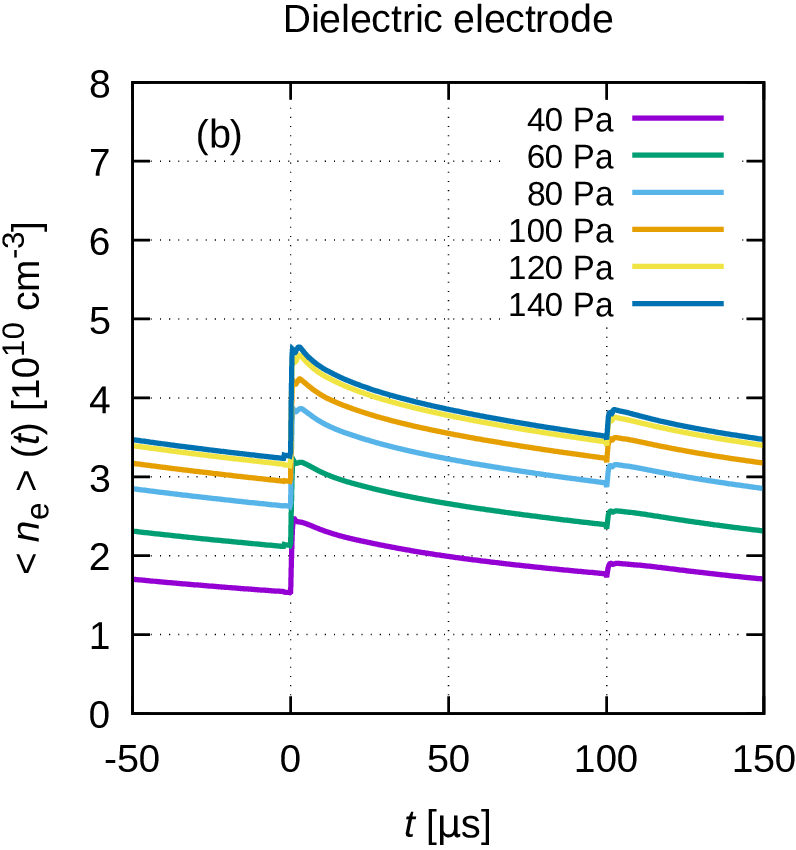}\\
\vspace{0.5cm}
~~~~\includegraphics[width=0.45\textwidth]{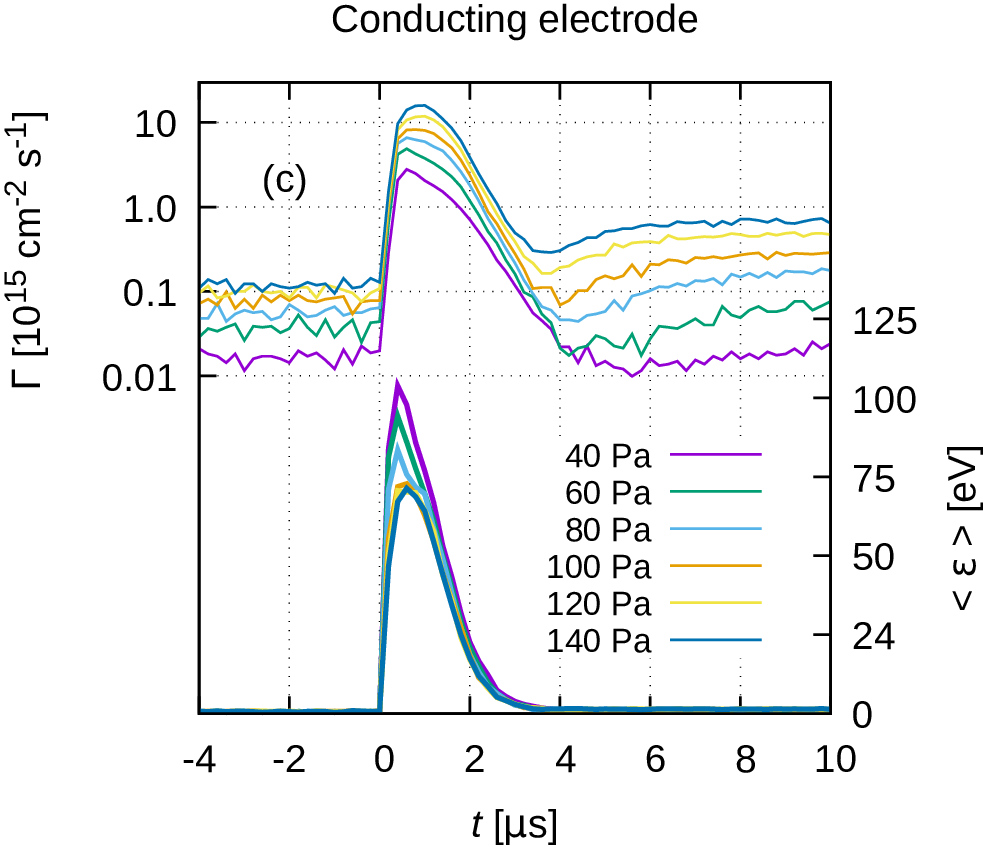}~~~~
\includegraphics[width=0.45\textwidth]{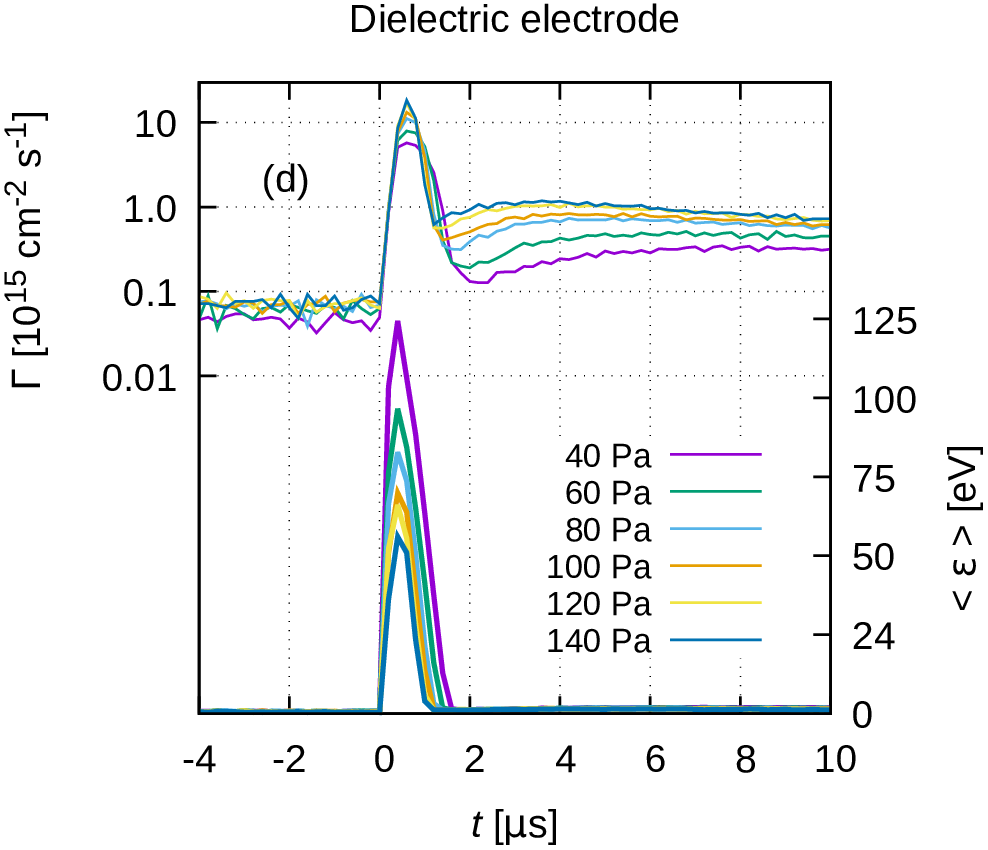}
\caption{Effect of the gas pressure on the discharge characteristics: (a,b) mean electron density as a function of time; (c,d) flux (left scale, upper set of thin lines) and mean energy (right scale, lower set of thick lines) of Ar$^+$ ions reaching the grounded electrode as a function of time. Results for the case of a conducting grounded electrode are shown in the left column, while results for the case of a dielectric electrode are shown in the right column. $\widehat{U}$ = 1000 V, $L$ = 3 cm, $\gamma = 0.07$ for the conducting surfaces and $\gamma = 0.3$ for the dielectric surface.}
\label{fig:par2}
\end{center}
\end{figure}

Figure~\ref{fig:par1} presents the simulation results for different driving voltage peak amplitudes, at fixed gas pressure and electrode separation ($p$ = 100 Pa and $L$ = 3 cm). Panels (a,c) and (b,d), respectively, refer to the cases of the conducting and dielectric grounded electrode. As it can be seen in panel (a), the electron density increases significantly with the voltage pulse amplitude. Doubled voltage pulse amplitude results in seven times increased electron density. The increase of the density upon the application of the excitation (i.e., the time-modulation of the mean electron density) is also more pronounced at higher voltage peak amplitudes due to more efficient electron multiplication in the avalanches near the temporary cathode.

For the grounded electrode covered by a dielectric layer, the electron density is generally lower as compared to the case of the conducting electrode (figure~\ref{fig:par1}(b)). The reason for this has already been discussed in section \ref{sec:res1}. The charging of the dielectric surface decreases the efficiency of the power coupling into the plasma due to the decrease of the gap voltage (cf. the discussion related to figure \ref{fig:voltages}), despite the higher secondary electron yield of the dielectric surface. We recall that the uneven increase of the density upon the application of the positive and negative pulses (at $t = 0\,\mu$s, and 100\,$\mu$s, respectively) is the consequence the different secondary electron emission coefficients adopted for conducting and dielectric surfaces. The mean ion energy during the application of the high-voltage pulses scales nearly linearly with the voltage amplitude, as it can be seen in figures~\ref{fig:par1}(c) and (d). The peak values of the ion flux are comparable in the cases of conducting and dielectric grounded electrode surfaces, and both exhibit an increase at higher voltages.

\begin{figure}[ht!]
\begin{center}
\includegraphics[width=0.44\textwidth]{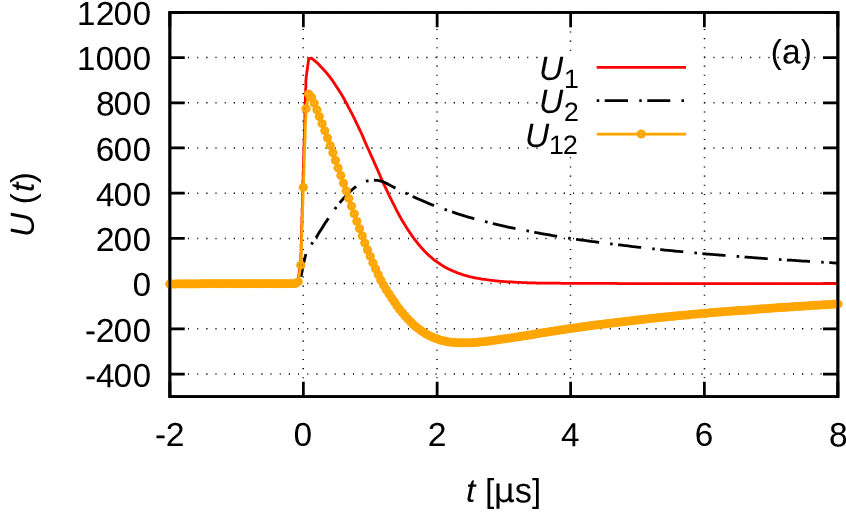}\\
\includegraphics[width=0.44\textwidth]{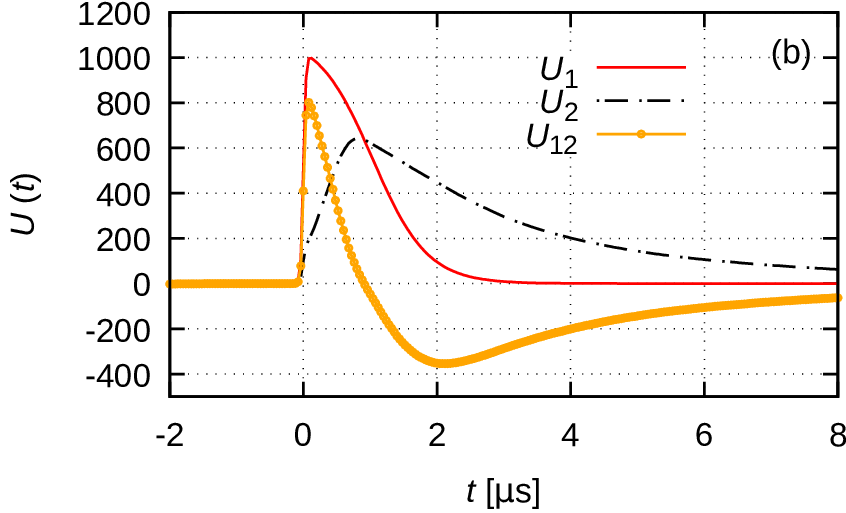}\\
\includegraphics[width=0.44\textwidth]{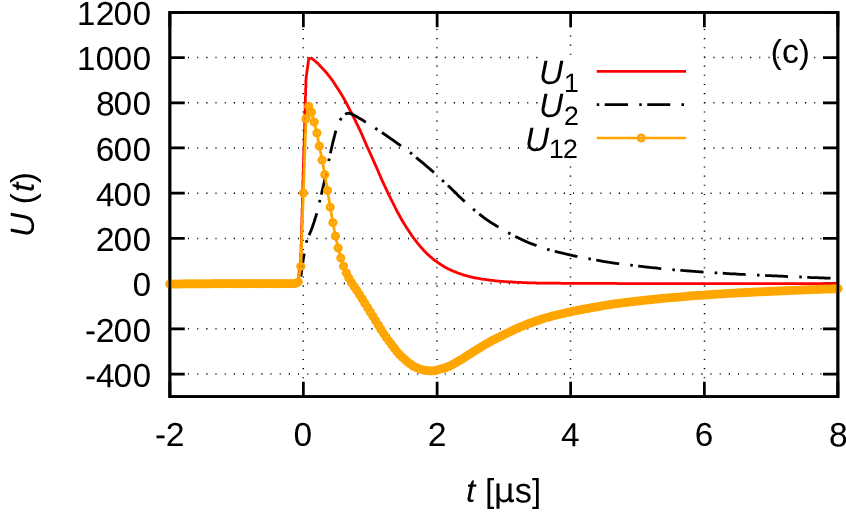}\\
\caption{Time dependence of the potentials of the electrodes. $U_1$ equals the applied voltage, $U_2$ is the surface potential of the dielectric at the grounded electrode, $U_{12} = U_1 - U_{2}$ is the discharge voltage. (a) 40 Pa, (b) 80 Pa, and (c) 140 Pa. $\widehat{U}_1 =$ 1000\,V, $L$ = 3\,cm.}
\label{fig:voltages_p}
\end{center}
\end{figure}

The effect of the gas pressure is revealed in figure \ref{fig:par2}. For the case of a conducting grounded electrode we observe a nearly linear increase of the mean electron density with increasing gas pressure (see figure \ref{fig:par2}(a)). In the case of a dielectric grounded electrode, a much less significant increase is observed (see figure \ref{fig:par2}(a)). It can be explained by the different "charging rates" of the dielectric surface of the grounded electrode as a function of the plasma density. We can follow the effect with the aid of figure \ref{fig:voltages_p} showing the time-dependence of the potentials of the electrodes and the gap voltage ($U_{12}$) for different gas pressures between 40 Pa and 140 Pa. While at the lowest pressure the potential at the dielectric surface rises up to $\approx$ 440\,V, at 80 Pa this value grows to 600\,V, and at 140 Pa to about 750\,V. This increase of the surface potential effectively decreases the gap voltage, and the shortening of the positive "wave" of $U_{12}$ with increasing pressure  becomes more significant. It limits the electron density that can be reached at a given voltage.

\begin{figure}[ht!]
\begin{center}
\includegraphics[width=0.36\textwidth]{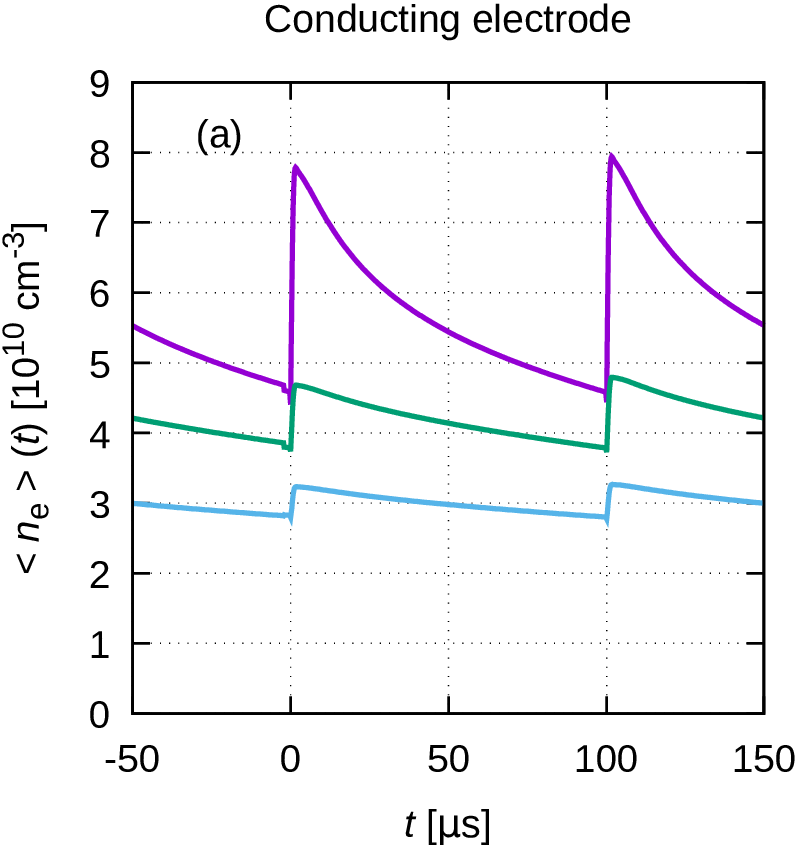}
~~~~~~~~~~~~~~~\includegraphics[width=0.36\textwidth]{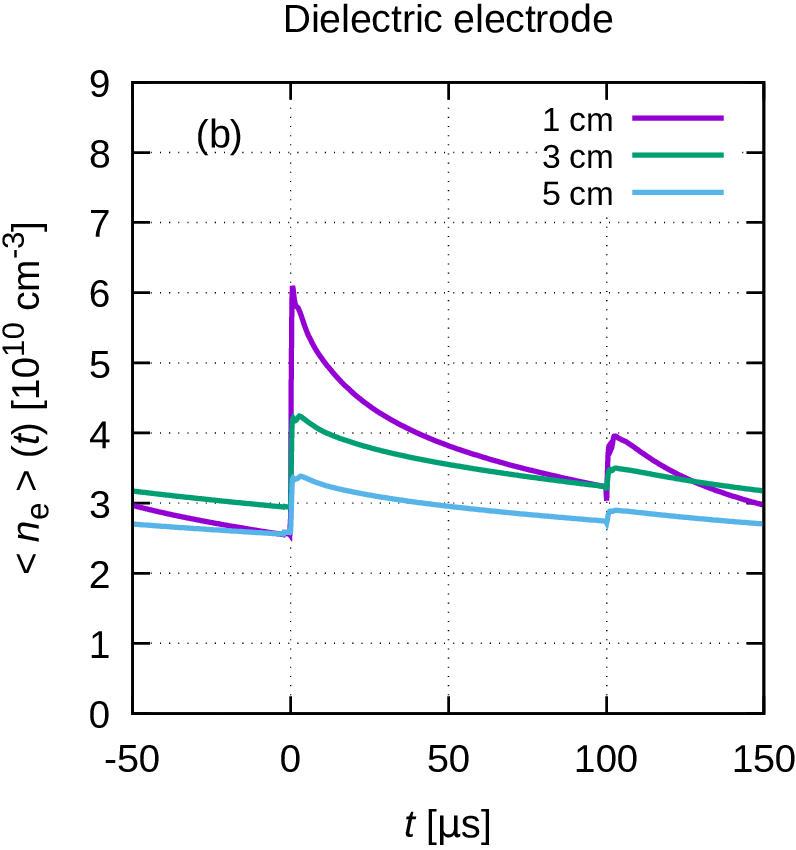}\\
\vspace{0.5cm}
~~~~\includegraphics[width=0.45\textwidth]{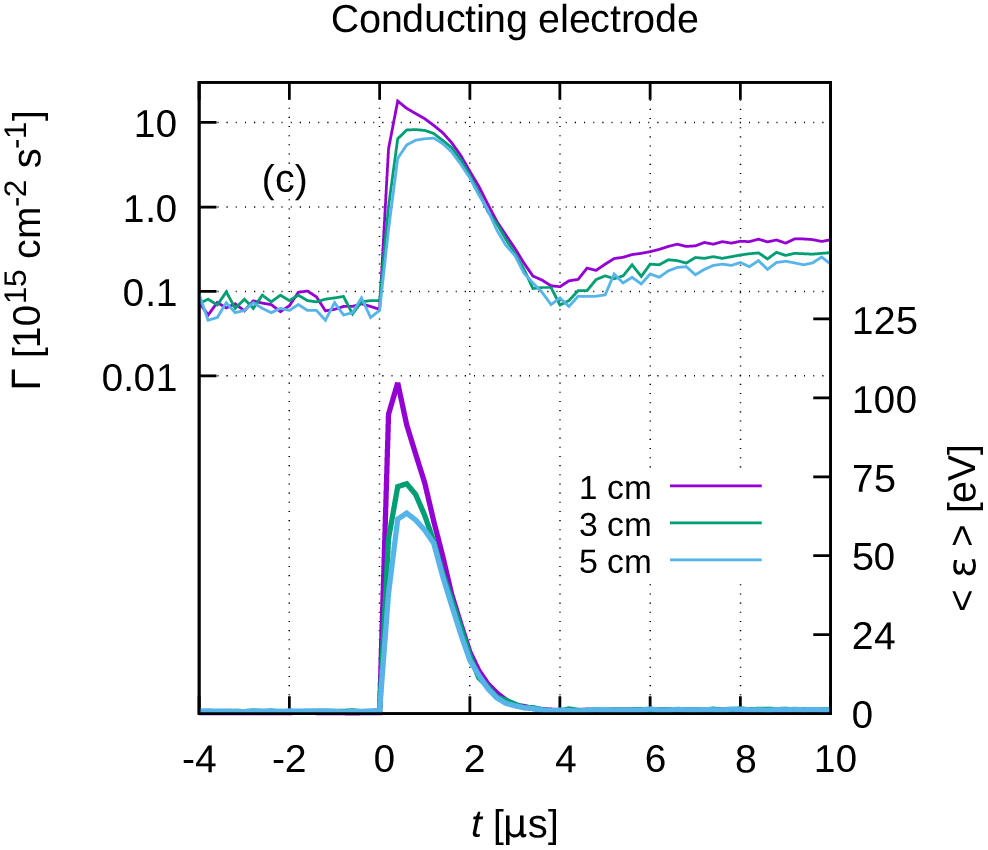}~~~~
\includegraphics[width=0.45\textwidth]{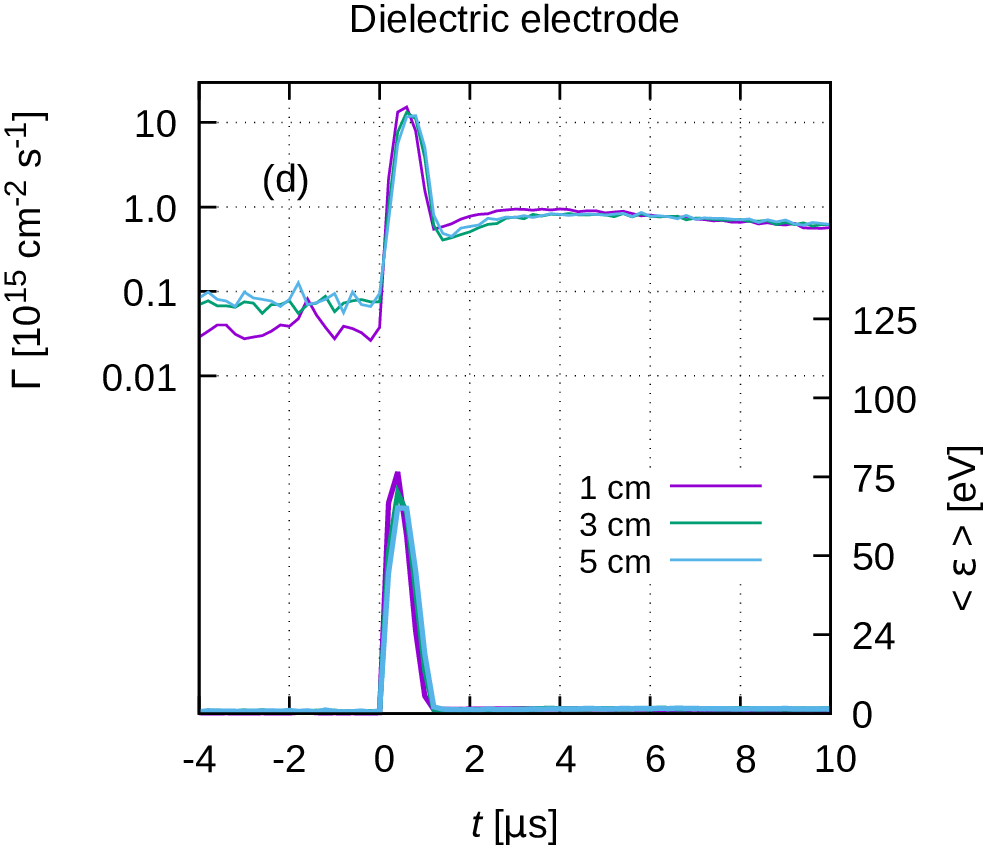}
\caption{Effect of the electrode gap on the discharge characteristics: (a,b) mean electron density as a function of time; (c,d) flux (left scale, upper set of thin lines) and mean energy (right scale, lower set of thick lines) of Ar$^+$ ions reaching the grounded electrode as a function of time. Results for the case of a conducting grounded electrode are shown in the left column, while results for the case of a dielectric electrode are in the right column. $\widehat{U}$ = 1000 V, $p$ = 100 Pa, $\gamma = 0.07$ for the conducting surfaces and $\gamma = 0.3$ for the dielectric surface.}
\label{fig:par3}
\end{center}
\end{figure}

The peak value of the ion flux at the grounded electrode increases nearly linearly with the gas pressure in the case of a conducting grounded electrode, following the behavior of the electron density (which is nearly the same as the ion density in the bulk plasma), see figures \ref{fig:par2}(a) and (c). In the case of a dielectric surface, the ion flux saturates with the increase of the pressure due to the reasons discussed above. Lower pressure result in higher ion energies due to a longer ion mean free path within the sheath region where a high electric field is present at times of high discharge voltage (see figures \ref{fig:par2}(b) and (d)).

The effects of the electrode gap ($L$) on the discharge characteristics (at a fixed voltage pulse amplitude of 1000\,V and a gas pressure of 100\,Pa) is demonstrated in figure \ref{fig:par3}. We observe a significant increase of the mean (spatially averaged) electron density with decreasing electrode gap. At a fixed pressure we expect that roughly the same number of electrons is created in the discharge pulses and the peak of the spatio-temporal distribution of the electrons is similar for different values of the gap. As the spatial average of the electron density at a larger $L$ is taken over a higher volume a lower $\langle n_{\rm e}\rangle$ is obtained at a bigger gap. Due to a faster diffusion decay a shorter gap results in a somewhat lower $\langle n_{\rm e}\rangle \, L$ product. 

\subsection{Comparison with experiments}

The simulation parameters employed in this study were inspired by recent plasma polymerization and surface treatment experiments carried out in high voltage bipolar-pulsed discharges \cite{Anjar19,Michlicek20}. However, no detailed plasma diagnostics have been performed for such plasmas yet. An earlier study \cite{Takechi01} presented the electron temperatures and densities obtained from Langmuir probe measurements at the center between the two electrodes in a similar parallel-plate high-voltage pulsed discharge system. In this experiment, the two parallel electrodes were made of Cu, i.\,e., both metallic, as sketched in figure 1(a). The electrodes had a diameter of 6 cm and  the gap between the electrodes was 3 cm. The discharge was ignited in Ar at the pressure of 100 Pa using mono-polar voltage pulses with peak values of $-350$ V $\sim$ $-650$ V  applied for a duration of 20 $\mu$s each. The interval between two consecutive pulses was 1 ms, i.\,e., the duty cycle was 1/50.

\begin{figure}[ht!]
\begin{center}
\includegraphics[width=0.5\textwidth]{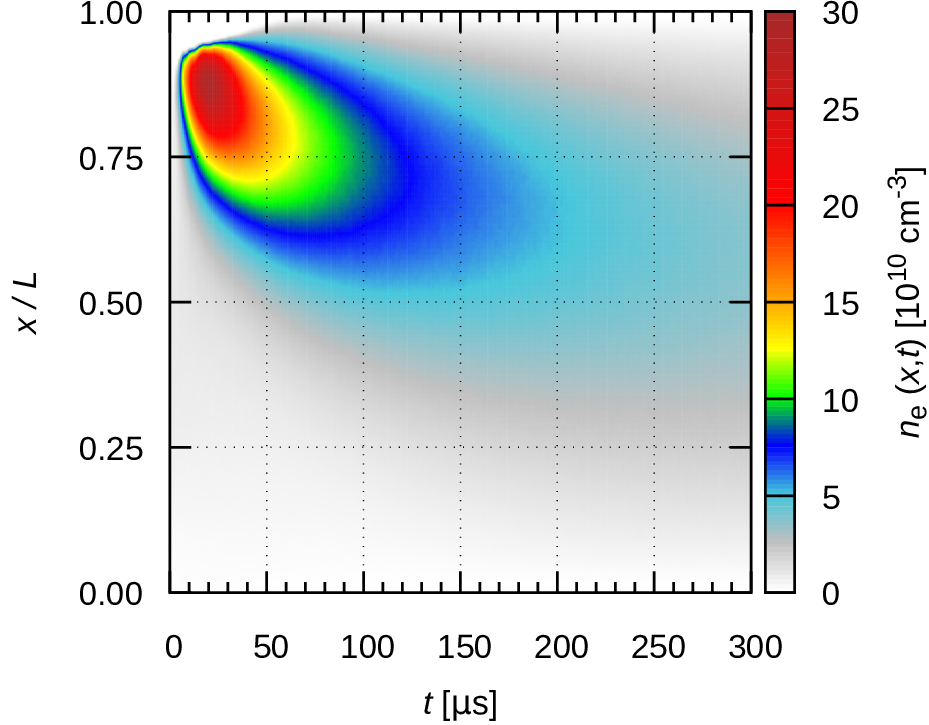}
\caption{Spatio-temporal distribution of the electron density for the case of an unipolar voltage pulse with $-600$\,V amplitude and 20 $\mu$s duration. The repetition rate is 1 ms
(the figure shows only a part of this period),  
$p$ = 100\,Pa, $L$ = 3\,cm, and $\gamma = 0.07$.
The metal grounded electrode is located at $x/L=0$, while the metal powered electrode is at $x/L=1$.}
\label{fig:experiment}
\end{center}
\end{figure}

For a comparison between the experiment and simulation, we take a $-600$ V pulse amplitude as a test case. For this case, the electron density at the middle of the electrode gap after the termination of the excitation pulses was found to be 4.5 $\times$ \, 10$^{10}$ cm$^{-3}$ in the experiment (see figure 3(b) of \cite{Takechi01}). Our numerical simulation under the same conditions (figure \ref{fig:experiment}) shows that the electron density at the center reaches a very similar value after the termination of the excitation pulse.  
However, in the experiment this density was measured at 5 $\mu$s after the termination of the excitation, while the simulation gives this density at somewhat later time (at about 50 $\mu$s). The different time-dependence that we observe may be due to uncertainties in the secondary electron emission coefficient and the pulse shapes in the experiment vs. the simulation, as well as due to the uncertainty of the probe measurements. The experiments and the simulations, on the other hand, both show (in agreement) that the central density hardly changes for a few hundred $\mu$s after the pulse.
The agreement between the experimental and simulation data shown here indicates that our simulations  presented in earlier subsections also reproduce corresponding discharge phenomena in experiments reasonably well.

\section{Summary}

\label{sec:summary} 

We have presented a numerical simulation study of a high voltage gas discharge, excited by microsecond pulses with alternating polarity and having a low duty cycle of approximately 1\%. Our studies have considered a parallel plate electrode configuration with a powered electrode made of conducting material and a grounded electrode with a conducting or a dielectric surface. The plasma source has been described using the Particle-in-Cell approach combined with the Monte Carlo treatment collision processes. 

A detailed analysis of the discharge was presented for a "base" set of conditions: 1000\,V voltage pulse amplitude, 100 Pa pressure and 3 cm electrode gap. The computed space- and time-resolved ionisation source function and electron density distribution provided detailed insight into the plasma formation. Tracing the ions in the simulations made it possible to derive the time dependence of their flux and the mean energy at the electrode surfaces. Despite the relatively high pressures considered here, it was found that the ion energies reach several hundreds of electron Volts. In the case of a dielectric-covered electrode, the plasma density and the ion flux to the electrodes were found to be self-limited as a function of the pressure, within the parameter range considered.

The energy distribution and the flux of the ions computed at numerous combinations of the external control parameters (voltage pulse amplitude, gas pressure and electrode gap) may aid the optimisation of the surface treatment applications of this type of plasma sources.

\ack This work was supported by the National Office for Research, Development and Innovation (NKFIH) of Hungary via the grant 119357, Osaka University International Joint Research Promotion Programs (Type A and Type B), Japan Society of the Promotion of Science (JSPS) Grant-in-Aid for Scientific Research(S) 15H05736, and JSPS Core-to-Core Program JPJSCCA2019002. A. H. is also grateful for the financial support by Research and Innovation in Science and Technology Project (RISET-PRO), Ministry of Research, Technology, and Higher Education of Indonesia. L.Z. is grateful to the Czech Science Foundation project 18-12774S and to the Ministry of Education, Youth and Sports of the Czech Republic supporting the project CEITEC 2020 (LQ1601).

\end{document}